\documentclass[10pt,journal,compsoc]{IEEEtran}
\ifCLASSOPTIONcompsoc
\usepackage[numbers,sort&compress]{natbib}
\usepackage{graphicx}
\usepackage{listings}
\usepackage{amssymb}
\usepackage[svgnames]{xcolor}
\usepackage{listofitems}
\usepackage{booktabs}
\usepackage{enumitem}
\usepackage{multirow}
\usepackage{eucal}
\usepackage{array}
\usepackage{color}
\usepackage{hyperref}
\usepackage{amsmath}
\usepackage{diagbox}
\newcommand\textlist[3][\cr]{%
  \readlist\indices{#3}%
  \foreachitem\x\in\indices{%
    \ifnum\xcnt=1\else, \fi$#2_{\x}$%
  }%
  \ifx\cr#1\relax\else%
    \if\relax#1\relax, \ldots\else, \ldots, $#2_{#1}$\fi%
  \fi%
}
\usepackage{subfigure}
\usepackage[normal]{caption}
\usepackage{amsmath}
\usepackage{algorithmic}
\begin{document}

\title{Combining Graph Neural Networks with Expert Knowledge for Smart Contract Vulnerability Detection}

\author{Zhenguang~Liu, 
        Peng~Qian, Xiaoyang Wang, Yuan~Zhuang, Lin~Qiu, and Xun~Wang 
       
\IEEEcompsocitemizethanks{
\IEEEcompsocthanksitem Zhenguang Liu, Peng Qian are with School of Computer and Information Engineering, Zhejiang Gongshang University and Zhejiang University, China. Email: liuzhenguang2008@gmail.com, messi.qp711@gmail.com
\IEEEcompsocthanksitem Yuan Zhuang is with National University of Singapore, Singapore.
\IEEEcompsocthanksitem Xiaoyang Wang is with School of Computer and Information Engineering, Zhejiang Gongshang University, China.
\IEEEcompsocthanksitem Lin Qiu is with Southern University of Science and Technology, China.
\IEEEcompsocthanksitem Xun Wang is with School of Computer and Information Engineering, Zhejiang Gongshang University and Zhejiang Lab, China. Email: xwang@zjgsu.edu.cn.
}
\thanks{Corresponding authors: Peng Qian, Xun Wang}
}
\markboth{IEEE Transactions on Knowledge and Data Engineering}
{Liu \MakeLowercase{\textit{et al.}}: Combining Expert Knowledge with Graph Neural Networks for Smart Contract Vulnerability Detection}

\IEEEtitleabstractindextext{
\begin{abstract}
Smart contract vulnerability detection draws extensive attention in recent years due to the substantial losses caused by hacker attacks. Existing efforts for contract security analysis heavily rely on rigid rules defined by experts, which are \emph{labor-intensive} and \emph{non-scalable}. More importantly, expert-defined rules tend to be \emph{error-prone} and suffer the inherent risk of being cheated by crafty attackers. Recent researches focus on the symbolic execution and formal analysis of smart contracts for vulnerability detection, yet to achieve a precise and scalable solution. Although several methods have been proposed to detect vulnerabilities in smart contracts, there is still a lack of effort that considers combining expert-defined security patterns with deep neural networks.

\quad In this paper, we explore using graph neural networks and expert knowledge for smart contract vulnerability detection. Specifically, we cast the rich control- and data- flow semantics of the source code into a \emph{contract graph}. To highlight the critical nodes in the graph, we further design a node elimination phase to normalize the graph. Then, we propose a novel temporal message propagation network to extract the graph feature from the normalized graph, and combine the graph feature with designed expert patterns to yield a final detection system. Extensive experiments are conducted on all the smart contracts that have source code in Ethereum and VNT Chain platforms. Empirical results show significant accuracy improvements over the state-of-the-art methods on three types of vulnerabilities, where the detection accuracy of our method reaches {89.15\%, 89.02\%, and 83.21\%} for reentrancy, timestamp dependence, and infinite loop vulnerabilities, respectively.
\end{abstract}

}

\maketitle

\IEEEdisplaynontitleabstractindextext
\IEEEpeerreviewmaketitle
\IEEEraisesectionheading{
\section{Introduction}
\label{sec:introduction}}
\IEEEPARstart{B}lockchain and its killer applications, \emph{e.g., Bitcoin and smart contract}, are taking the world by storm \cite{blockbench,Yaga,bitcoin,Dhawan,tsikhanovich2019pd,dinh2018untangling}. A blockchain is essentially a distributed and shared transaction ledger, maintained by all the miners in the blockchain network following a consensus protocol \cite{consensus}. The consensus protocol and replicated ledgers enforce all the transactions immutable once recorded on the chain, endowing blockchain with  decentralization and tamper-free nature.

\textbf{Smart contract.} Smart contracts are programs running on top of the blockchain \cite{Dhawan,oyente}. A smart contract can implement arbitrary rules for managing assets by encoding the rules into source code. The defined rules of a contract will be strictly and automatically followed during execution, effectuating the `code is law' logic. Smart contracts make the automatic execution of contract terms possible, facilitating complex decentralized applications (DApps). Indeed, many DApps are basically composed of several smart contracts as the backend and a user interface as the frontend \cite{dapps}.

Millions of smart contracts have been deployed in various blockchain platforms, enabling a wide range of applications including wallets \cite{wallet}, crowdfunding, decentralized gambling \cite{decentralized}, and cross-industry finance \cite{industry}. The number of smart contracts is still growing rapidly. For example, within the last six months, over 15,000 new contracts were deployed on Ethereum alone, which is the most famous smart contract platform. 

\textbf{Security issues of smart contracts.} Smart contracts from various fields now hold more than 10 billion dollars worth of virtual coins. Undoubtedly, holding so much wealth makes smart contracts attractive enough to attackers. In June 2016, attackers exploited the reentrancy vulnerability of the DAO contract \cite{DAO} to steal 3.6 million Ether, which was worth 60 million US Dollars. This case is not isolated and several security vulnerabilities are discovered and exploited every few months \cite{DAO,King,Multisig}, undermining the trust for smart contract-based applications.

There are several reasons that make smart contracts particularly prone to errors. \emph{First}, the programming languages (\emph{e.g., Solidity}) and tools are still new and crude, leaving plenty of rooms for bugs and misunderstandings in the tools \cite{oyente,smartcheck}. \emph{Second}, since smart contracts are immutable once deployed, developers are required to anticipate all possible status and environments the contract may encounter in the future, which is undoubtedly difficult. Distinct from conventional distributed applications that can be updated when bugs are detected, there is no way to patch the bugs of a smart contract without forking the blockchain (almost an impossible task), regardless of how much money the contract holds or how popular it is \cite{oyente}. Therefore, effective vulnerability checkers for smart contracts before their deployment are essential.

\textbf{Drawbacks of conventional methods.} Conventional methods for smart contract vulnerability detection, \emph{such as} \cite{oyente,smartcheck,contractfuzzer,securify}, employ classical static analysis or dynamic execution techniques to identify vulnerabilities. Unfortunately, they fundamentally rely on several expert-defined patterns. The manually defined patterns bear the inherent risk of being \emph{error-prone} and some complex patterns are \emph{non-trivial} to be covered. Crudely using several rigid patterns leads to high \emph{false-positive} and \emph{false-negative} rates, and crafty attackers may easily bypass the pattern checking using tricks. Moreover, as the number of smart contracts increases rapidly, it is becoming impossible for a few experts to sift through all the contracts to design precise patterns. A feasible solution might be: ask each expert to label a number of contracts, then collect all the labeled contracts from many experts to train a model that can automatically give a prediction on whether a contract has a specific type of vulnerability. 

Recently, efforts have been made towards adopting deep neural networks for smart contract vulnerability detection \cite{pengqian,ijcai,Wesley}, achieving improved accuracy. \cite{pengqian} utilizes LSTM based networks to sequentially process source code, while \cite{ijcai} models the source code into control flow graphs. \cite{Wesley} builds a sequential model to analyze the Ethereum operation code. However, these approaches either treat the source code or operation code as a text sequence instead of semantic blocks, or fail to highlight critical variables in the data flow \cite{ijcai}, leading to insufficient semantic modeling and unsatisfactory results. 

To fill the research gap, in this paper, we investigate more than 300,000 smart contract functions and present a fully automated and scalable approach that can detect vulnerabilities at the function level. Specifically, we cast the rich control- and data- flow semantics of the source code into graphs. The nodes in the graph represent critical variables and function invocations, while directed edges capture their temporal execution traces. Since not all nodes in the graph are of equal importance and most graph neural networks are inherently flat during information propagation on the graph, we design a node elimination phase to normalize the graph and highlight the key nodes. The normalized graph is then fed into a temporal message propagation network to learn the \emph{graph} feature. In the meantime, we extract the \emph{security pattern} feature from the source code using expert knowledge. Finally, the \emph{graph} feature and \emph{security pattern} feature are incorporated to produce the final vulnerability detection results. 

We conducted experiments on all the 40k contracts that have source code in Ethereum and on all the contracts in VNT Chain, demonstrating significant improvements over state-of-the-art vulnerability detection methods: F1 score from $78\%$ to $86\%$, $79\%$ to $88\%$, $74\%$ to $82\%$ for \emph{reentrancy, timestamp dependence}, and \emph{infinite loop} vulnerabilities, respectively. Our implementations\footnote{Github: https://github.com/Messi-Q/GPSCVulDetector} are released to facilitate future research.

We would like to point out that this work is clearly distinct from the previous one \cite{ijcai} in three ways: 1) this work is to investigate whether combining graph neural networks with conventional expert patterns could achieve better vulnerability detection results, while the objective of the previous work is to explore the possibility of using neural networks for smart contract vulnerability detection. 2) In this work, we propose to extract vulnerability-specific expert patterns and combine them with the graph feature. We also explicitly model the key variables in the data flow. In contrast, in the previous work, we only utilize the graph feature while ignoring expert patterns and key variables. 3) This work consistently outperforms the previous one across different vulnerabilities, and overall provides more insights and findings in this field. Note that in the previous paper, we proposed two neural networks, \emph{DR-GCN} and \emph{TMP}, to explore the applicability of different graph neural networks on smart contract vulnerability detection. In this paper, we focus on extending TMP, which delivers better performance than \emph{DR-GCN}. We will also extend \emph{DR-GCN} and compare it with the extension of TMP. 
 
\textbf{Contributions.} To summarize, the key contributions are: 

\begin{itemize} 
\item
To the best of our knowledge, we are the first to investigate the idea of fusing conventional expert patterns and graph-neural-network extracted features for smart contract vulnerability detection. 
\item We propose to characterize the contract function source code as contract graphs. We also explicitly normalize the graph to highlight key variables and invocations. A novel temporal message propagation network is proposed to automatically capture semantic graph features.  
\item Our methods set the new state-of-the-art performance on smart contract vulnerability detection, and overall provide insights into the challenges and opportunities. As a side contribution, we have released our implementations to facilitate future research.
\end{itemize}

\vspace{-0.7em}
\section{Related Work}
\label{related_work}

\subsection{Smart Contract Vulnerability Detection}
Smart contract vulnerability detection is one of the fundamental problems in blockchain security. Early works on smart contract vulnerability detection verify smart contracts by employing formal methods \cite{Bhargavan,Grishchenko,Hildenbrandt,Hirai}. For example, \cite{Bhargavan} introduces a framework, translating \emph{Solidity} code (the smart contract programming language of Ethereum) and the EVM (Ethereum Virtual Machine) bytecode into the input of an existing verification system. \cite{Hirai} proposes a formal model for EVM and reasons the potential bugs in smart contracts by using the Isabelle/HOL tool. Further, \cite{Grishchenko} and \cite{Hildenbrandt} define formal semantics of the EVM using the F* framework and the $\mathbb{K}$ framework, respectively. Although these frameworks provide strong formal verification guarantees, they are still semi-automated. 

Another stream of work relies on generic testing and symbolic execution, such as Oyente \cite{oyente}, Maian \cite{Maian}, and Securify \cite{securify}. Oyente is one of the pioneering works that perform symbolic execution on contract functions and flags bugs based on simple patterns. Zeus \cite{zeus} leverages abstract interpretation and symbolic model checking, as well as the constrained horn clauses to detect vulnerabilities in smart contracts. \cite{securify} introduces compliance (negative) and violation (positive) patterns to filter false warnings. 

Researchers also explore smart contract vulnerability detection using dynamic execution. \cite{contractfuzzer} presents ContractFuzzer to identify vulnerabilities by fuzzing and runtime behavior monitoring during execution. Similarly, \cite{reguard} develops a fuzzing-based analyzer to identify the reentrancy vulnerability. Sereum \cite{sereum} uses taint analysis to monitor runtime data flows during smart contract execution for vulnerability detection. However, dynamic execution methods require a hand-crafted agent contract to interact with the contract under test, preventing them from fully-automated applications and endowing them non-scalability.  

Recently, a few attempts have been made to study using deep neural networks for smart contract vulnerability detection. \cite{pengqian} constructs the sequential \emph{contract snippet} and feeds them into the BLSTM-ATT model to detect reentrancy bugs. \cite{ijcai} proposes to convert the source code of a contract into the \emph{contract graph} and constructs graph neural networks as the detection model. \cite{wang2020contractward} proposes ContractWard, extracting bigram features from the operation code of smart contracts and utilizing machine learning algorithms. However, although a few methods have been proposed, the field of contract vulnerability detection using deep learning is still in its infancy and the accuracy is still unsatisfactory. For common smart contract vulnerabilities and attacks, motivated readers may refer to \cite{Sok} for a comprehensive survey.

\vspace{-0.7em}
\subsection{Graph Neural Network}
With the remarkable success of neural networks, graph neural network has been investigated extensively in various fields such as graph classification \cite{Zhang,wang2016incremental}, program analysis \cite{devign,Miltiadis}, and graph embedding \cite{cai2018comprehensive}. Existing approaches roughly cast into two categories: \textbf{(i)} \emph{Spectral-based approaches} generalize well-established neural networks like CNNs to work on graph-structured data. For instance, GCN \cite{GCN} implements a first-order approximation of spectral graph convolutions \cite{Defferrard,zhou2018graph,wei2019mmgcn} and develops a layer-wise propagation network using the Laplacian matrix, which achieves promising performance on graph node classification tasks. \cite{adaptive} proposes a graph CNN which can take data of arbitrary graph structure as input. \textbf{(ii)} \emph{Spatial-based methods} inherit ideas from recurrent GNNs and adopt information propagation to define graph convolutions. Early work such as \cite{Micheli} directly sums up the nodes' neighborhood information for graph convolutions. Another line of work, such as GAT \cite{velivckovic2017graph} and GAAN \cite{zhang2018gaan}, employs attention mechanisms to learn the weights of different neighboring nodes. Motivated by these spatial-based approaches, \cite{gilmer2017neural} outlines a message-passing neural network framework to predict the chemical properties of molecules.

\begin{figure}
    \begin{center}
     \includegraphics[width=3.1in]{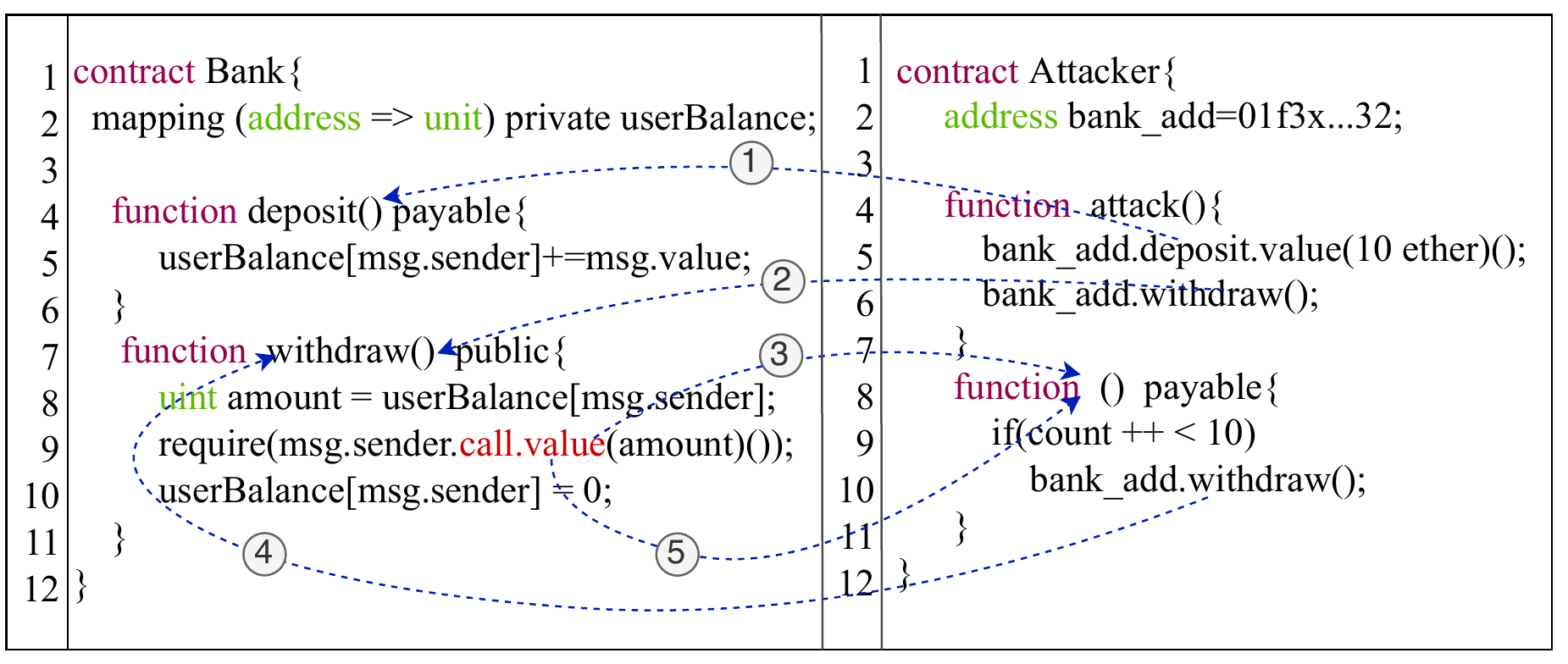}
    \end{center}
     \vspace{-1.0em}
 \caption{A simplified example of reentrancy vulnerability.}
  \label{fig_dao}
  \vspace{-1.5em}
\end{figure}

Recently, \cite{Miltiadis,xu2017neural,devign,shen2019neuro} attempt to apply GNNs to program analysis issues. Specifically, \cite{Miltiadis} introduces a gated graph recurrent network for variable prediction, while \cite{xu2017neural} proposes Gemini for binary code similarity detection, where functions in binary code are represented by attributed control flow graphs. \cite{devign} develops Devign, a general graph neural network-based model for vulnerability identification in C programming language. Different from these methods, we focus on the specific smart contract vulnerability task, and explicitly take into account the distinct roles and temporal relationships of program elements.

\begin{figure*}
    \begin{center}
     \includegraphics[width=7.1in]{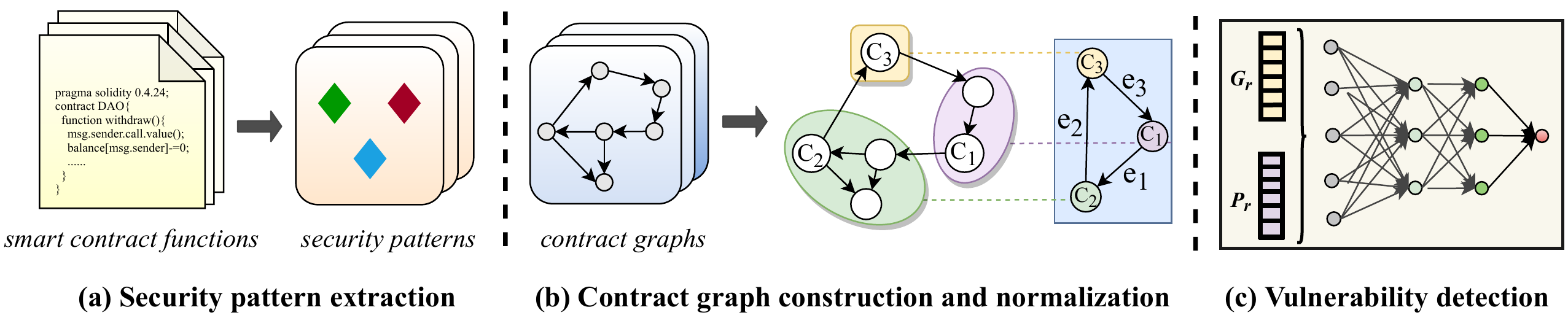}
\vspace{-1.0em}
    \end{center}
 \caption{The overall architecture of our proposed method. (a) The expert pattern extraction phase; (b) the contract graph construction and normalization phase; (c) the vulnerability detection phase.} 
 \label{fig_framework}
 \vspace{-1.6em}
\end{figure*}

\section{Problem Statement}
\label{sec_problem}
In this section, we first formulate the problem, then introduce the three types of vulnerabilities studied in this paper, and present the reasons for focusing on these three vulnerabilities.

\label{preliminaries} 
\textbf{Problem formulation.} Given the source code of a smart contract, we are interested in developing a fully automated approach that can detect vulnerabilities at the function level. In other words, we are to estimate the label $\hat{y}$ for each smart contract function $f$, where $\hat{y}$ = 1 represents $f$ has a specific vulnerability while $\hat{y} = 0$ denotes $f$ is safe. In this paper, we focus on three types of vulnerabilities, which will be presented below. Before that, we first introduce the preliminary knowledge of the fallback mechanism in smart contracts, which is important in understanding the problem.

\textbf{Fallback mechanism.} Within a smart contract, each function is uniquely identified by a signature, consisting of its name and parameter types \cite{Sok}. Upon a function invocation, the signature of the invoked function is passed to the called contract. If the signature matches a function of the called contract, the execution jumps to the corresponding function. Otherwise, it jumps to the fallback function. Money transfer is considered as an empty signature, which will trigger the fallback function as well. The fallback function is a special function with no name and no argument, which can be arbitrarily programmed \cite{Sok}. After introducing this background knowledge, we now are ready to elaborate on the three types of vulnerabilities.

(1) \textbf{Reentrancy} is a well-known vulnerability that caused the infamous DAO attack. When a smart contract function $f_1$ transfers money to a recipient contract $C$, the fallback function $f_2$ of $C$ will be automatically executed. In its fallback function $f_2$, $C$ may invoke back to $f_1$  for conducting an invalid second-time transfer. Since the current execution of $f_1$ waits for the first-time transfer to finish, $C$ can make use of the intermediate state of $f_1$ to succeed in stealing money. A simplified example is shown in Fig. \ref{fig_dao}, where the \emph{withdraw} function of contract \emph{Bank} has a reentrancy vulnerability, contract \emph{Attacker} steals money by exploiting the vulnerability. First, \emph{Attacker} deposits 10 Ether (Ether is the virtual money of Ethereum) in contract \emph{Bank} (step 1). Then, \emph{Attacker} withdraws the 10 Ether by invoking the {\emph{withdraw}} function (step 2). When the contract \emph{Bank} sends 10 Ether to \emph{Attacker} using \textbf{\emph{call.value}} (\emph{Bank}, line 9), the fallback function (\emph{Attacker}, lines 8--11) of \emph{Attacker} will be automatically invoked (step 3). In its fallback function, \emph{Attacker} calls \textbf{\emph{withdraw}} again (step 4). Since the {\emph{userBalance}} of \emph{Attacker} has not yet been set to 0 (\emph{Bank}, line 10), \emph{Bank} believes that \emph{Attacker} still has {10 Ether} in the contract, thus transfers 10 Ether to \emph{Attacker} again (Step 5). The withdraw loop lasts for 9 times ({\small$count++ < 10$}, \emph{Attacker} line 9). Finally, \emph{Attacker} obtains much more Ether (100 Ether) than expected (10 Ether).

(2) \textbf{Timestamp dependence} vulnerability exists when a smart contract uses the block timestamp as a triggering condition to execute some critical operations, e.g., using the \emph{timestamp} of a future block as the source to generate random numbers so as to determine the winner of a game. The miner (a node in the blockchain) who mines the block has the freedom to set the timestamp of the block within a short time interval ($<$ 900 seconds) \cite{contractfuzzer}. Therefore, miners may manipulate the block timestamps to gain illegal benefits.

(3) \textbf{Infinite loop} is a common vulnerability in smart contracts. The program of a function may contain a loop (e.g. \emph{for} loop, \emph{while} loop, and self-invocation loop) with no exit condition or the exit condition cannot be reached, namely an infinite loop. 

\textbf{Why focus on these vulnerabilities.} We mainly focus on the three aforementioned vulnerabilities since: \textbf{(i)} In real attacks, blockchain networks have suffered more than 100 million USD losses due to the three vulnerabilities. For instance, attacks on reentrancy have caused one of the biggest losses (60 million USD in The Dao Event) in smart contract history. \textbf{(ii)} We empirically found that the three vulnerabilities may affect a significant number of smart contracts and are non-trivial to be detected. Specifically, we surveyed 40,932 Ethereum smart contracts, observing that around 5,013 out of 307,396 functions  possess at least one invocation to \emph{call.value}. Although possessing a \emph{call.value} invocation does not necessarily mean that the contract has a reentrancy vulnerability, the contract has the potential to be affected by the \emph{reentrancy} vulnerability and thus requires further checking. Similarly, around 4,833 functions have used \emph{block.timestamp} and thus are potentially affected by the \emph{timestamp dependence} vulnerability. Many functions have \emph{for} or \emph{while} loops, which may lead to the \emph{infinite loop} vulnerability. In contrast, most other contract vulnerabilities affect a relatively smaller number of functions, \emph{e.g.}, the \emph{locked ether} vulnerability affects less than 900 functions, and the \emph{integer overflow} vulnerability affects less than 1,000 functions.

\vspace{-0.7em}
\section{Our Method}
\label{our_method}
\textbf{Method overview.} The overall architecture of our proposed method is depicted in Fig. \ref{fig_framework}, which consists of three phases: (1) a security pattern extraction phase, which obtains the vulnerability-specific expert patterns from the source code; (2) a contract graph construction and normalization phase, which extracts the control flow and data flow semantics from the source code and highlights the critical nodes; and (3) a vulnerability detection phase, which casts the normalized contract graph into graph feature using temporal graph neural network, and combines the pattern feature and graph feature to output the detection result. In what follows, we elaborate on the details of the three components one by one.
\begin{figure*}
	\centering
	\includegraphics[width=7.1in]{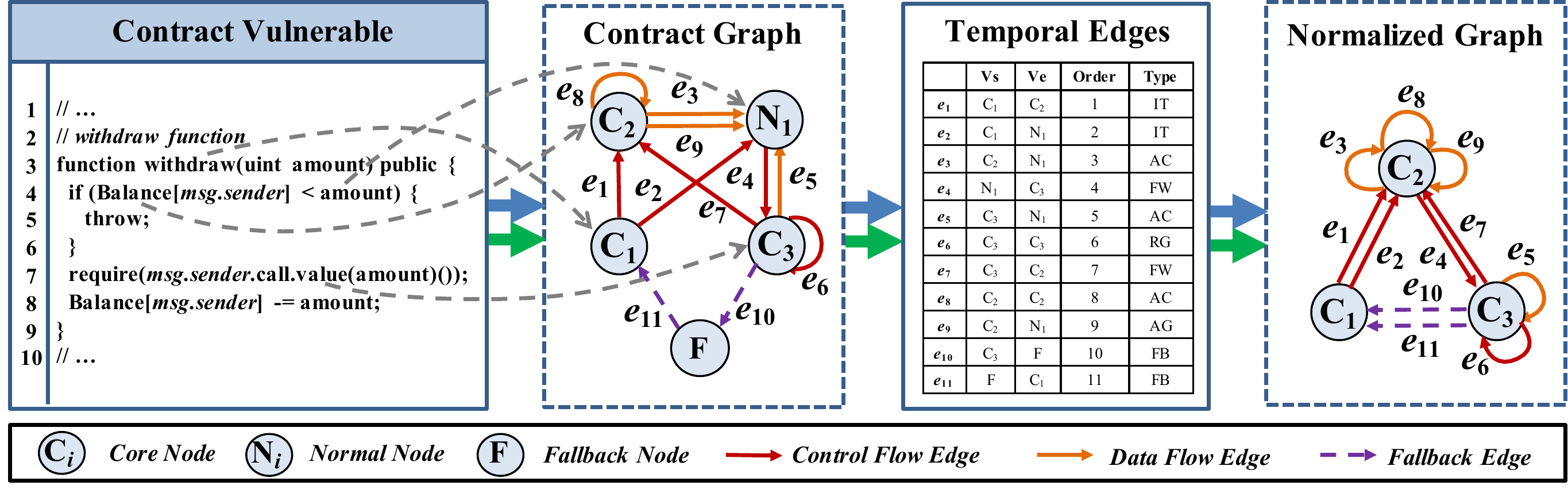} 
	\caption{The contract graph construction and normalization phase. The first figure shows the source code of a contract function, while the second figure visualizes the contract graph extracted from the code. Nodes $\bold{C_i}$ denote core nodes, nodes $\bold{N_i}$ represent normal nodes, and node $\bold{F}$ denotes fallback node. The third figure illustrates the temporal edges in the extracted graph, where the types of edges are detailed in Table \ref{symbols}. The fourth figure demonstrates the graph after normalization.}
	\label{fig:normalization}
	 \vspace{-1.6em}
\end{figure*}

\vspace{-0.7em}
\subsection{Expert Pattern Extraction}
\label{pattern_representation}
In this section, we summarize existing patterns and design new patterns for the three specific vulnerabilities respectively, and implement an open-sourced tool to automatically extract these patterns.

\textbf{Reentrancy.} Conventionally, the reentrancy vulnerability is considered as an invocation to \emph{call.value} that can call back to itself through a chain of calls. That is, the invocation of \emph{call.value} is successfully re-entered to perform the unexpected operation of repeated money transfer. By investigating existing works such as \cite{oyente,contractfuzzer,zeus}, we design three sub-patterns. The first sub-pattern is \textbf{callValueInvocation} that checks whether there exists an invocation to \emph{call.value} in the function. The second sub-pattern \textbf{balanceDeduction} checks whether the user balance is deducted \emph{after} money transfer using \emph{call.value}, which considers the fact that the money stealing can be avoided if user balance is deducted each time \emph{before} money transfer. The third sub-pattern \textbf{enoughBalance} concerns whether there is a check on the sufficiency of the user balance before transferring to a user. Note that \emph{enoughBalance} is a new pattern designed in this paper.

\textbf{Timestamp dependence.} Generally, the timestamp dependence vulnerability exists when a smart contract uses the block timestamp as part of the conditions to perform critical operations \cite{contractfuzzer}. By investigating previous works including \cite{Sok,oyente,contractfuzzer}, we design three sub-patterns that are closely related to timestamp dependence. \emph{First,} sub-pattern \textbf{timestampInvocation} models whether there exists  an invocation to opcode \emph{block.timestamp} in the function. \emph{Then,} the second sub-pattern \textbf{timestampAssign} checks whether the value of \emph{block.timestamp} is assigned to other variables or passed to a function as a parameter, namely whether \emph{block.timestamp} is actually used. \emph{Last,} the third sub-pattern \textbf{timestampContamination} checks if \emph{block.timestamp} may contaminate the triggering condition of a critical operation, which can be implemented by taint analysis. Sub-pattern \emph{timestampContamination} is a new pattern designed in this paper.

\textbf{Infinite loop.} Infinite loop is conventionally considered as a loop bug which unintentionally iterates forever, failing to jump out of the loop and return an expected result. Specifically, we define three expert patterns for infinite loop as follows. (1) The first sub-pattern \textbf{loopStatement} checks whether the function possesses a loop statement such as \emph{for} and \emph{while}. (2) The second sub-pattern \textbf{loopCondition} models whether the exit condition can be reached. For example, for a \emph{while} loop, its exit condition $i<10$ may not be reached if $i$ is never updated in the loop. (3) The third sub-pattern \textbf{selfInvocation} models whether the function invokes itself and the invocation is not in an \emph{if} statement. This concerns the fact that if the self-invocation statement is not in an \emph{if} statement, the self-invocation loop will never terminate.

\textbf{Pattern Extraction Implementations.} We implemented an open-sourced tool to extract the designed expert patterns from smart contract functions. Particularly, simple sub-patterns such as \emph{callValueInvocation}, \emph{timestampInvocation}, and \emph{loopStatement} can be directly extracted by keyword matching. Sub-patterns \emph{balanceDeduction}, \emph{enoughBalance}, \emph{loopCondition}, \emph{timestampAssign}, and \emph{selfInvocation} are obtained by syntax analysis. Complex sub-pattern \emph{timestampContamination} is extracted by taint analysis where we follow the traces of the data flow and flag all the variables that may be affected along the traces.

\vspace{-0.7em}
\subsection{Contract Graph Construction and Normalization}
\label{graph_representation}
Existing works \cite{Miltiadis,rossi2018deep} have shown that programs can be transformed into symbolic graph representations, which are able to preserve semantic relationships (e.g., data dependency and control dependency) between program elements. Inspired by this, we formulate smart contract functions into \emph{contract graphs}, and assign distinct roles to different program elements (namely nodes). We also construct edges to model control and data flow between program elements, taking their temporal orders into consideration. Further, we design a node elimination process to normalize the \emph{contract graph} and highlight important nodes. Next, we introduce contract graph construction and normalization, respectively.

\vspace{-0.7em}
\subsubsection{Contract Graph Construction}
\label{nodes_and_edges}
\textbf{Nodes construction.} Our first insight is that different program elements in a function are not of equal importance in detecting vulnerabilities. Therefore, we extract three types of nodes, \emph{i.e.}, \textit{core nodes}, \textit{normal nodes}, and \textit{fallback nodes}.

\textit{Core nodes.} Core nodes symbolize the key invocations and variables that are critical for detecting a specific vulnerability. In particular, for reentrancy vulnerability, core nodes model (i) an invocation to a money transfer function or the built-in \emph{call.value} function, (ii) the  variable that corresponds to \emph{user balance}, and (iii) variables that can directly affect \emph{user balance}. For timestamp dependence vulnerability,  (i) invocations to \textit{block.timestamp}, (ii) variables assigned by \textit{block.timestamp}, and (iii) invocations to a random function that takes \textit{block.timestamp} as the cardinal seed are extracted as core nodes. For infinite loop vulnerability, (i) all the loop statements such as \emph{for} and \emph{while} statements, (ii) the loop condition variables, and (iii)  self invocations are considered as core nodes.

\textit{Normal nodes.} While core nodes represent key invocations and variables, normal nodes are used to model invocations and variables that play an auxiliary role in detecting vulnerabilities. Specifically, invocations and variables that are not extracted as core nodes are modeled as normal ones, e.g., for timestamp dependence vulnerability, invocations that do not call \emph{block.timestamp} and variables indirectly related to \emph{block.timestamp} are considered as normal nodes.  

\textit{Fallback node.} Further, we construct a fallback node $F$ to stimulate the fallback function of a {virtual} attack contract, which can interact with the function under test. 

\emph{A simplified example.} Taking contract \emph{Vulnerable} presented in the left of Fig. \ref{fig:normalization} as an example, suppose we are to evaluate whether its \emph{withdraw} function possesses a reentrancy vulnerability. As shown by the arrows in the left two figures of Fig. \ref{fig:normalization}, function \emph{withdraw} itself is first modeled as a core node $C_1$ since its inner code contains \emph{call.value}. Then, following the temporal order of the code, we treat the critical variable $Balance[msg.sender]$ as a core node $C_2$, while variable $amount$ is modeled as normal node $N_1$. The invocation to \emph{call.value} is extracted as a core node $C_3$, and the \emph{fallback} function of a virtual attack contract is characterized by the fallback node $F$.

\begin{table}
\centering
\resizebox{0.4\textwidth}{!}{
\begin{tabular}{|c|c|c|}
\hline
\multicolumn{1}{|c|}{\textbf{Type (Abbreviation)}} & \textbf{Semantic Fact} & \textbf{Category} \\
\hline
AH & assert\{X\} & \multirow{11}{*}{\begin{tabular}[c]{@{}c@{}}Control-flow\end{tabular}} \\
RG & require\{X\} & \\
IR & if\{...\} revert & \\
IT & if\{...\} throw & \\
IF & if\{X\} & \\
GB & if\{...\} else \{X\} & \\
GN & if\{...\} then \{X\} & \\
WH & while\( \{ X\}\) do\{...\} & \\
FR & for\( \{ X\}\) do\{...\} & \\
FW & natural sequential relationships & \\
\hline
\multicolumn{1}{|c|}{AG} & assign\{X\} & \multirow{2}{*}{\begin{tabular}[c]{@{}c@{}}Data-flow\end{tabular}} \\
\multicolumn{1}{|c|}{AC} & access\{X\} & \\
\hline
\multicolumn{1}{|c|}{FB} & interactions with fallback function& Fallback \\
\hline
\end{tabular}
}
\caption{Semantic edges summarization. All edges are classified into three categories, namely control-flow, data-flow, and fallback edges.}
\label{symbols}
 \vspace{-1.8em}
\end{table}

\textbf{Edges construction.} Our second insight is that the nodes are closely related to each other in a temporal manner rather than being isolated. To capture rich semantic dependencies between the nodes, we construct three categories of edges, namely \emph{control flow, data flow}, and \emph{fallback} edges. Each edge describes a path that might be traversed through by the function under test, and the temporal number of the edge characterizes its sequential order in the function. We investigated various functions and summarized the semantic edges in Table \ref{symbols}. All edges are classified into three categories. 

\emph{Control flow edges.} Control flow edges capture the control semantics of the code. Specifically, a control flow edge is constructed for a \emph{conditional} statement or \emph{security handle} statement, such as a \emph{if, for, assert}, and \emph{require} statement. The edge directs from the previous node encountered, which represents the critical function call or variable preceding to the current statement, to the node representing the function call or variable in the current statement. In particular, we use forward edges to describe the natural control flow of the code sequence. A forward edge connects two nodes in the adjacent statements. The main benefit of such encoding is to reserve the programming logic reflected by the sequence of the source code. The control flow edges are depicted with red arrows in Fig. \ref{fig:normalization}.

\emph{Data flow edges.} Data flow edges track the usage of variables. A data flow edge involves the access or modification of a variable. The data flow edges are demonstrated with orange arrows in Fig. \ref{fig:normalization}. For example, the \emph{access} and \emph{assign} statement $Balance[msg.sender]$-=$amount$ (line 8, \emph{Vulnerable}, Fig. \ref{fig:normalization}) is characterized by two data flow edges, i.e., an access edge $e_7$ starting from the $Balance[msg.sender]$ variable node $C_2$ to itself, and an assign edge $e_8$ starting from $C_2$ to the \emph{amount} variable node $N_1$.

\emph{Fallback edges.} {In order to explicitly model the specific fallback mechanism, two fallback edges are constructed. The first fallback edge connects from the first \emph{call.value} invocation to the fallback node, while the second edge directs from the fallback node to the function under test. The fallback edges are shown by dashed purple edges in Fig. \ref{fig:normalization}}.

\begin{figure}
    \begin{center}
     \includegraphics[width=3.2in]{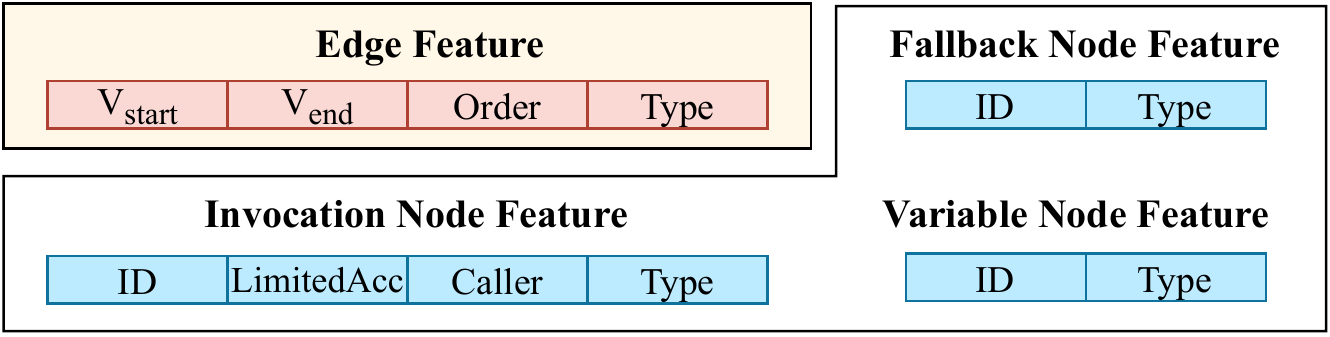}
    \end{center}
 \caption{Illustration of the edge feature and node feature.}
  \label{fig:features}
   \vspace{-1.6em}
\end{figure}

\textbf{Node and edge features.} Fig. \ref{fig:features} illustrates the extracted features for edges and nodes, respectively. Specifically, the feature of an edge is extracted as a tuple ($V_{start}$, $V_{end}$, \emph{Order}, \emph{Type}), where $V_{start}$ and $V_{end}$ represent its start and end nodes, \emph{Order} denotes its temporal order, and \emph{Type} stands for edge type. For nodes, different kinds of nodes possess different features. 1) The feature of a node that models function invocation consists of (\emph{ID}, \emph{AccFlag}, \emph{Caller}, \emph{Type}), where \emph{ID} denotes its identifier, \emph{Caller} represents the caller address of the invocation, and $Type$ stands for the node type. Interestingly, the modifier of a smart contract function $\Psi$ may trigger the pre-check of certain conditions, e.g., modifier \emph{owner} will check whether the caller of $\Psi$ is the owner of the contract before executing $\Psi$. Therefore, we use \emph{AccFlag} to capture this semantics, where \emph{AccFlag} = `LimitedACC' specifies the function has limited access while AccFlag =`NoLimited' denotes non-limited access. 2) In contrast, the feature of a fallback node or a node that models variable consists of only \emph{ID} and \emph{Type}.

\begin{figure*}
    \begin{center}
     \includegraphics[width=5.8in]{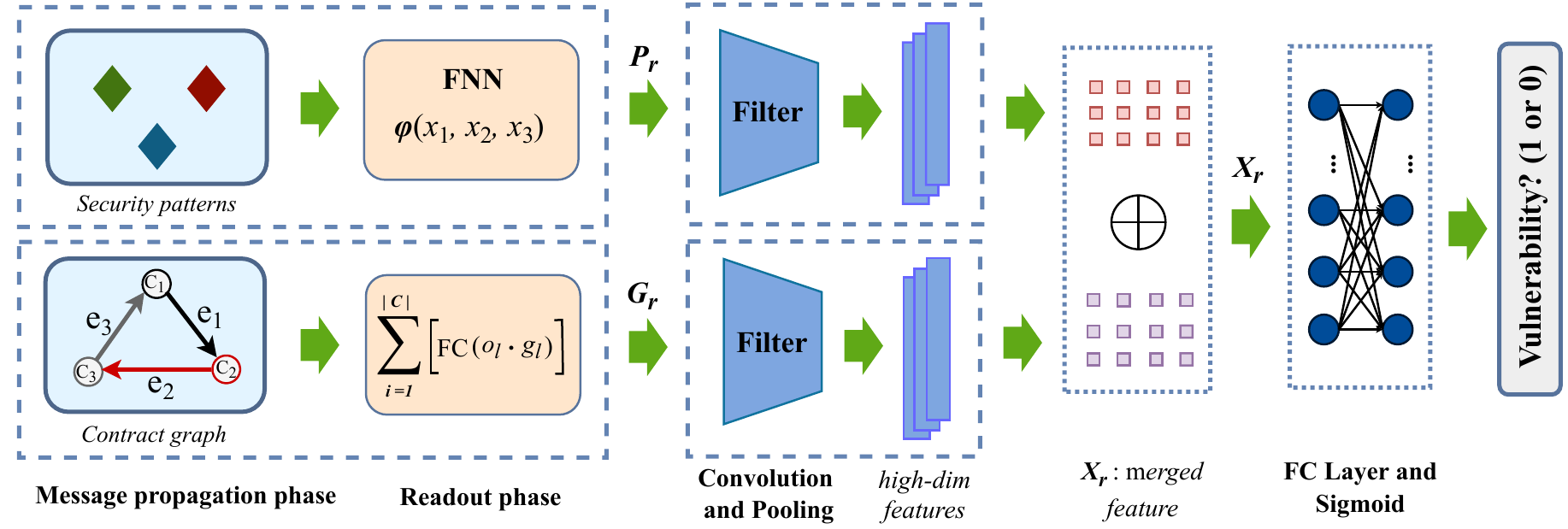}
    \end{center}
 \caption{The process of vulnerability detection. First, a feed-forward neural network generates the pattern feature $P_{r}$ for the security patterns extracted from the source code. Then, the temporal message propagation network  is used to extract the graph feature $G_{r}$ from the contract graph. Finally, the CGE network combines $G_{r}$ and $P_{r}$ into the merged feature $X_{r}$, which is fed into the FC and sigmoid layers to output the vulnerability detection results.}
  \label{fig:classifier}
   \vspace{-1.6em}
\end{figure*}

\subsubsection{Contract Graph Normalization}
\label{graph_normalization}
Most graph neural networks are inherently flat when propagating information, ignoring that some nodes play more central roles than others. Moreover, different contract functions yield distinct graphs, hindering the training of graph neural networks. Therefore, we propose a node elimination process to normalize the \emph{contract graph}.

\textbf{Nodes elimination.} As introduced in Section~\ref{nodes_and_edges}, the nodes of a \emph{contract graph} are partitioned into core nodes $\{C_{i}\}_{i=1}^{|C|}$, normal nodes $\{N_{i}\}_{i=1}^{|N|}$, and the fallback node $F$. We remove each normal node $N_i$, but pass the feature of $N_i$ to its nearest core nodes. For example, the normal node $N_1$ in the second figure of Fig.~\ref{fig:normalization} is removed with its feature aggregated to nearest core nodes $C_2$ and $C_3$. For a node $N_i$ that has multiple nearest core nodes, its feature is passed to all of them. The edges connected to the removed normal nodes are preserved but with their start or end node moving to the corresponding core node. The fallback node is also removed similar to the normal node. 

\textbf{Feature aggregation.} After removing normal nodes, features of core nodes are updated by aggregating features from their neighboring normal nodes. More precisely, the new feature of $C_i$ is composed of three components: \textbf{(i)} self-feature, namely the feature of core node $C_i$ itself; \textbf{(ii)} in-features, namely features of the normal nodes $\{P_{j}\}^{|P|}_{j=1}$ that are merged to $C_i$ and having a path pointing from $P_{j}$ to $C_i$; and \textbf{(iii)} out-feature, namely features of the normal nodes $\{Q_{k}\}^{|Q|}_{k=1}$ that are merged to $C_i$ and having a path directs from $Q_{k}$ to $C_i$. Note that features of different normal nodes that model variables and invocations are added respectively when aggregating to the same node.

\vspace{-0.7em}
\subsection{Vulnerability Detection} 
\label{vulnerability_detection}
In this subsection, we introduce the proposed vulnerability detection network CGE (\underline{C}ombining \underline{G}raph feature and \underline{E}xpert patterns). First, we obtain the expert pattern feature $P_{r}$ by passing the extracted sub-patterns (introduced in subsection~\ref{pattern_representation}) into a feed-forward neural network (FNN). Then, we extract the graph feature $G_{r}$ from the normalized \emph{contract graph} by our proposed temporal message propagation network, consisting of a \emph{message propagation} phase and a \emph{readout} phase. Finally, we use a fusion network to combine the graph feature $G_{r}$ and the pattern feature $P_{r}$, outputting the detection results. The process is demonstrated in Fig. ~\ref{fig:classifier} with details presented below.

\textbf{Security pattern feature $P_{r}$ extraction.} For the sub-patterns closely related to a specific vulnerability, we utilize a one-hot vector to represent each sub-pattern, and append a 0/1 digit to each vector, which indicates whether the function under test has the sub-pattern. The vectors for all sub-patterns related to a specific vulnerability are concatenated into a final vector $x$. Taking $x$ as the input, and the ground truth  of whether the function has the specific vulnerability as the target label, we utilize a feed-forward neural network $\varphi (x) $ to extract high-dimensional semantic feature $P_{r} \in \mathbb{R}^{d}$. 

\textbf{Contract graph feature $G_r$ Extraction}. 
After extracting security pattern feature $P_{r}$, we further obtain the semantic feature of the contract graph by using our proposed temporal-message-propagation network, which consists of a \emph{message propagation} phase and a \emph{readout} phase. In the message propagation phase, the network passes information along the edges successively by following their temporal orders. Then, it generates the graph feature $G_{r}$ by using a readout function, which aggregates the final states of all nodes in the contract graph.

\emph{Message propagation phase.} Formally, we denote the normalized contract graph as $G = \{V, E\}$, where $V$ consists of the core nodes, and $E$ consists of all edges. Denote $E$ = \{\textlist[N]{e}{1,2}\}, where $e_{k}$ represents the $k^{th}$ temporal edge.

Messages are passed along the edges, one edge per time step. At first, the hidden state $h^{0}_{i}$ for each node $V_{i}$ is initialized with its own node feature. Then, at time step $k$, message flows through the $k^{th}$ temporal edge $e_{k}$ and updates the hidden state $h_{ek}$ of the end node of $e_{k}$. 

More specifically, message $m_{k}$ is first computed basing on the hidden state $h_{sk}$ of the start node of $e_{k}$, and the edge type $t_{k}$:
\begin{align}
\label{eq:e1}
x_{k} = h_{sk} \oplus t_{k} \\
m_{k} = W_{k}x_{k} + b_{k}
\end{align}
where $\oplus$ denotes concatenation, matrix $W_{k}$ and bias vector $b_k$ are network parameters. The original message $x_{k}$ contains information from the start node of $e_{k}$ and edge $e_{k}$ itself, which are then transformed into a vector embedding using $W_{k}$ and $b_k$.

After receiving the message, the end node of $e_{k}$ updates its hidden state $h_{ek}$ by aggregating information from the incoming message and its previous state. Formally, $h_{ek}$ is updated according to:
\begin{align}
\label{eq:e2}
\hat{h}_{ek} = tanh(Um_{k} + Zh_{ek} + b_{1}) \\
h^{'}_{ek} = softmax(R\hat{h}_{ek} + b_{2})
\end{align}
where network parameters $U$, $Z$, $R$ are matrices, while $b_{1}$ and $b_{2}$ are bias vectors.

\emph{Readout phase.} After successively traversing all the edges in $G$, we extract the feature for $G$ by reading out the final hidden states of all nodes. Let $h^{T}_{i}$ be the final hidden state of the $i^{th}$ node, we find that the differences between the final hidden state $h^{T}_{i}$ and the original hidden state $h^{0}_{i}$ are informative in the vulnerability detection task. Therefore, we consider to generate the graph  feature $G_{r}$ by
\begin{align}
\label{eq:e3}
s_{i} &= h^{T}_{i}\oplus h^{0}_{i} \\
g_{i} &= softmax(W^{(2)}_{g}(tanh(b^{(1)}_{g} + W^{(1)}_{g}s_{i})) + b^{(2)}_{g}) \\
o_{i} &= softmax(W^{(2)}_{o}(tanh(b^{(1)}_{o} + W^{(1)}_{o}s_{i})) + b^{(2)}_{o}) \\
G_{r} &= FC(\sum^{|V|}_{i = 1}o_{i} \odot g_{i})
\end{align}
where $\oplus$ denotes concatenation, and $\odot$ denotes element-wise product. $W_{j}$, $b^{(1)}_{j}$, and $b^{(2)}_{j}$, with subscript $j \in \{g, o\}$ are network parameters.

\textbf{Vulnerability detection by combining $P_r$ and $G_r$.} After obtaining the security pattern feature $P_{r}$ and the contract graph feature $G_{r}$, we combine them to compute the final label $\hat{y} \in (0, 1)$, indicating whether the function under test has the specific vulnerability. To this end, we first filter $P_{r}$ and $G_{r}$ using a convolution layer and a max pooling layer, then we concatenate the filtered features and pass them to a network consisting of 3 fully connected layers and a sigmoid layer. The process can be formulated as: 
\begin{align}
\label{eq:e4}
X_{r} = \psi(P_{r}) \oplus \psi(G_{r}) \\
\hat{y} = sigmoid(FC(X_{r}))
\end{align}
The convolutional layer learns to assign different weights to different elements of the semantic vector, while the max pooling layer highlights the significant elements and avoids overfitting. The fully connected layer and the non-linear sigmoid layer produce the final estimated label $\hat{y}$.

\vspace{-0.7em}
\section{Experiments}
\label{experiments}
In this section, we empirically evaluate our proposed methods on all the Ethereum smart contracts that have source code verified by Etherscan \cite{Etherscan}, as well as on all the available smart contracts on another blockchain platform VNT Chain \cite{Vntchain}. We seek to answer the following research questions:
\begin{itemize}[noitemsep,wide=0pt, leftmargin=\dimexpr\labelwidth + 2\labelsep\relax]
\item \textbf{RQ1}: Can the proposed method effectively detect the reentrancy, infinite loop, and timestamp dependence vulnerabilities? How are its \emph{accuracy}, \emph{precision}, \emph{recall}, and \emph{F1 score} performance against the state-of-the-art conventional vulnerability detection approaches?
\item \textbf{RQ2}: Is our proposed method useful for detecting new types of vulnerabilities, e.g., sharing-variable reentrancy, which is difficult for existing methods?
\item \textbf{RQ3}: Can the proposed method outperform other neural network-based methods? 
\item \textbf{RQ4}: How do the proposed \emph{security pattern}, \emph{graph normalization}, \emph{message propagation} modules, and \emph{different network layers} in \emph{CGE} affect the performance of the proposed method?
\end{itemize}
Next, we first present the experimental settings, followed by answering the above research questions one by one.

\renewcommand\arraystretch{1.0}
\begin{table*}
\centering
\resizebox{0.999\textwidth}{!}{
\begin{tabular}{cccccccccccccc}
\toprule
\multirow{2}{*}{\textbf{Methods}} & \multicolumn{4}{c}{\textbf{Reentrancy (ESC dataset)}} & \multicolumn{4}{c}{\textbf{Timestamp dependence (ESC dataset)}} & \multirow{2}{*}{\textbf{Methods}} & \multicolumn{4}{c}{\textbf{Infinite Loop (VSC dataset)}}\\
\cmidrule(lr){2-5}\cmidrule(lr){6-9}\cmidrule(lr){11-14} & Acc(\%) & Recall(\%) & Precision(\%) & F1(\%) & Acc(\%) & Recall(\%) & Precision(\%) & F1(\%) & & Acc(\%) & Recall(\%) & Precision(\%) & F1(\%) \\
\midrule
Smartcheck & 52.97 & 32.08 & 25.00 & 28.10 & 44.32 & 37.25 & 39.16 & 38.18 & Jolt & 42.88 & 23.11 & 38.23 & 28.81 \\
Oyente & 61.62 & 54.71 & 38.16 & 44.96 & 59.45 & 38.44 & 45.16 & 41.53 & PDA & 46.44 & 21.73 & 42.96 & 28.26 \\
Mythril & 60.54 & 71.69 & 39.58 & 51.02 & 61.08 & 41.72 & 50.00 & 45.49 & SMT & 54.04 & 39.23 & 55.69 & 45.98 \\
Securify & 71.89 & 56.60 & 50.85 & 53.57 & -- & -- & -- & -- & Looper & 59.56 & 47.21 & 62.72 & 53.87 \\
Slither & 77.12 & 74.28 & 68.42 & 71.23 & 74.20 & 72.38 & 67.25 & 69.72 & -- & -- & -- & -- & -- \\
\midrule
Vanilla-RNN & 49.64 & 58.78 & 49.82 & 50.71 & 49.77 & 44.59 & 51.91 & 45.62 & Vanilla-RNN & 49.57 & 47.86 & 42.10 & 44.79 \\
LSTM & 53.68 & 67.82 & 51.65 & 58.64 & 50.79 & 59.23 & 50.32 & 54.41 & LSTM & 51.28 & 57.26 & 44.07 & 49.80 \\
GRU & 54.54 & 71.30 & 53.10 & 60.87 & 52.06 & 59.91 & 49.41 & 54.15 & GRU & 51.70 & 50.42 & 45.00 & 47.55 \\
\textbf{GCN} & \textbf{77.85} & \textbf{78.79} & \textbf{70.02} & \textbf{74.15} & \textbf{74.21} & \textbf{75.97} & \textbf{68.35} & \textbf{71.96} & \textbf{GCN} & \textbf{64.01} & \textbf{63.04} & \textbf{59.96} & \textbf{61.46} \\	
\midrule	
{DR-GCN} & {81.47} & {80.89} & {72.36} & {76.39} & {78.68} & {78.91} & {71.29} & {74.91} & {DR-GCN} & {68.34} & {67.82} & {64.89} & {66.32} \\
\textbf{TMP} & \textbf{84.48} & \textbf{82.63} & \textbf{74.06} & \textbf{78.11} & \textbf{83.45} & \textbf{83.82} & \textbf{75.05} & \textbf{79.19} & \textbf{TMP} & \textbf{74.61} & \textbf{74.32} & \textbf{73.89} & \textbf{74.10} \\
\textbf{CGE} & \textbf{89.15} & \textbf{87.62} & \textbf{85.24} & \textbf{86.41} & \textbf{89.02} &  \textbf{88.10} & \textbf{87.41} & \textbf{87.75} & \textbf{CGE} & \textbf{83.21} & \textbf{82.29} & \textbf{81.97} & \textbf{82.13} \\
\bottomrule
\end{tabular}
}
\caption{ Performance comparison in terms of \emph{accuracy}, \emph{recall}, \emph{precision}, and \emph{F1 score}. A total of sixteen methods are investigated in the comparison, including state-of-the-art vulnerability detection methods, neural network-based alternatives, DR-GCN, TMP, and CGE. ‘--’ denotes not applicable.}
\label{Performance_comparison}
 \vspace{-1.8em}
\end{table*}

\vspace{-0.7em}
\subsection{Experimental Settings}
\textbf{Datasets.} We conducted experiments on two real-world smart contract datasets, namely \texttt{ESC} (Ethereum Smart Contracts) and \texttt{VSC} (VNT chain Smart Contracts), which are collected from Ethereum and VNT Chain platforms, respectively. Experiments for reentrancy and timestamp dependence vulnerabilities are conducted on \texttt{ESC}, while the infinite loop vulnerability is evaluated on \texttt{VSC}.
\begin{itemize}
\item The \texttt{ESC} dataset consists of  307,396 smart contract functions from  40,932 smart contracts in Ethereum \cite{Ethereum}. Among the functions, around 5,013 functions possess at least one invocation to \emph{call.value}, making them potentially affected by the reentrancy vulnerability. Around 4,833 functions contain the \emph{block.timestamp} statement, making them susceptible to the timestamp dependence vulnerability.  {Around 56,800 functions contain \emph{for} or \emph{while} loop statements}. 
\item The \texttt{VSC} dataset contains $13,761$ functions, which are collected from all the available $4,170$ smart contracts in the VNT Chain network \cite{Vntchain}. VNT Chain is an experimental public blockchain platform proposed by companies and universities from Singapore, China, and Australia.  The VNT Chain runs smart contracts written in Class C language.
\end{itemize}

\textbf{Implementation details.} All the experiments are conducted on a computer equipped with an Intel Core i7 CPU at 3.7GHz, a GPU at 1080Ti, and 32GB of Memory. Our vulnerability detection system consists of three main components: the auto \emph{CodeExtractor} tool for extracting the security patterns and contract graphs from the source code; the \emph{Normalization} tool for normalizing contract graphs; the CGE network that outputs  results by combining pattern feature and graph feature. The \emph{CodeExtractor} and \emph{Normalization} tools are implemented with Python, while the CGE network is implemented with TensorFlow. The implementations of our vulnerability detection system are available at \url{https://github.com/Messi-Q/GPSCVulDetector}. 

\textbf{Parameter settings.} The adam optimizer is employed in the CGE network. We apply a grid search to find out the best settings of hyper-parameters: the learning rate $l$ is tuned amongst \{0.0001, 0.0005, 0.001, 0.002, 0.005, 0.01\}, the dropout rate $d$ is searched in \{0.1, 0.2, 0.3, 0.4, 0.5\}, and batch size $\beta$ in \{8, 16, 32, 64, 128\}. To prevent overfitting, we tuned the L2 regularization $\lambda$ in \{$10^{-6}$, $10^{-5}$, $10^{-4}$, $10^{-3}$, $10^{-2}$, $10^{-1}$\}. Without special mention in texts, we report the performance of all neural network models with following default setting: 1) $l=0.002$, 2) $d=0.2$, 3) $\beta=32$, and 4) $\lambda=10^{-4}$. For each dataset, we randomly select $80\%$ of them as the training set and the other $20\%$ as the testing set for several times, and report the averaged result. The ground truth labels for contract functions are provided by experts.

\vspace{-0.7em}
\subsection{Comparison with  State-of-the-art Existing Methods (RQ1)}
\label{comparisons}
In this section, we benchmark our proposed method against existing non-deep-learning vulnerability detection approaches, which include:
\begin{itemize}[noitemsep,wide=0pt, leftmargin=\dimexpr\labelwidth + 2\labelsep\relax]
	\item \textbf{Oyente} \cite{oyente}: A well-known symbolic verification tool for smart contract vulnerability detection, which performs symbolic execution on the CFG (control flow graph) to check vulnerable patterns.
	\item \textbf{Mythril} \cite{mythril}: A security analysis method, which uses concolic analysis, taint analysis, and control flow checking to detect smart contract vulnerabilities.
	\item \textbf{Smartcheck} \cite{smartcheck}: An extensible static analysis tool for discovering smart contract code vulnerabilities.
	\item \textbf{Securify} \cite{securify}: A formal-verification based tool for detecting Ethereum smart contract bugs, which checks compliance and violation patterns to filter false positives.
	\item \textbf{Slither} \cite{feist2019slither}: A static analysis framework designed to find issues in Ethereum smart contracts by converting a smart contract into an intermediate representation of \emph{SlithIR}.
\end{itemize}

\textbf{Comparison on reentrancy vulnerability detection.} First, we compare our CGE approach with the five existing methods on the reentrancy vulnerability detection task. The performance of different methods is presented in the left of Table~\ref{Performance_comparison}, where metrics of \emph{accuracy}, \emph{recall}, \emph{precision}, and \emph{F1 score} are engaged. We would like to highlight that all metrics are computed over only the susceptible smart contract functions having invocation(s) to \emph{call.value}, i.e., the functions that may be infected with the reentrancy vulnerability. Functions with no \emph{call.value} invocation are known to be immune to reentrancy vulnerability and is trivial to be handled (using purely keyword matching), thus we do not involve those functions in the calculation to better investigate the problem. From the quantitative results of Table \ref{Performance_comparison}, we have the following observations. First, we find that conventional non-deep-learning methods have not yet achieved a satisfactory accuracy on the reentrancy vulnerability detection task, e.g., the state-of-the-art method (i.e., Slither) yields a $77.12\%$ accuracy. Second, our proposed method substantially outperforms the existing methods on reentrancy vulnerability detection. Specifically, CGE achieves a 89.15\% accuracy, gaining a 12.03\% accuracy improvement over conventional methods. The strong empirical evidences suggest the great potential of combing graph neural networks with expert patterns for reentrancy vulnerability detection. 

By looking into the existing methods, we believe that the reasons for the low precision and recall of conventional methods are: (1) they heavily rely on simple and fixed patterns to detect vulnerabilities, \emph{e.g., Mythril checks whether the call.value invocation is not followed by any internal function call to detect reentrancy}, and (2) the rich data dependencies and control dependencies within smart contract code are not characterized with fine-grained details in these methods. 

\textbf{Comparison on timestamp dependence vulnerability detection.} We further compare the proposed CGE with the five methods on the timestamp dependence vulnerability detection task. The comparison results are demonstrated in the middle of Table~\ref{Performance_comparison}. The state-of-the-art conventional method (i.e., Slither) has obtained a 74.20\% accuracy on timestamp dependence vulnerability detection, which is quite low. This may stem from the fact that most of existing methods detect timestamp dependence vulnerability by crudely checking whether there is \emph{block.timestamp} statement in the function. Moreover, in consistent with the results on reentrancy vulnerability detection, CGE keeps delivering the best performance in terms of all the four metrics. In particular, CGE gains a 14.82\% accuracy improvement over state-of-the-art conventional methods.

\begin{figure*}
\centering 
\resizebox{0.205\textwidth}{!}{
\subfigure[Reentrancy comparison of tools]{
    \includegraphics[width=4.5cm]{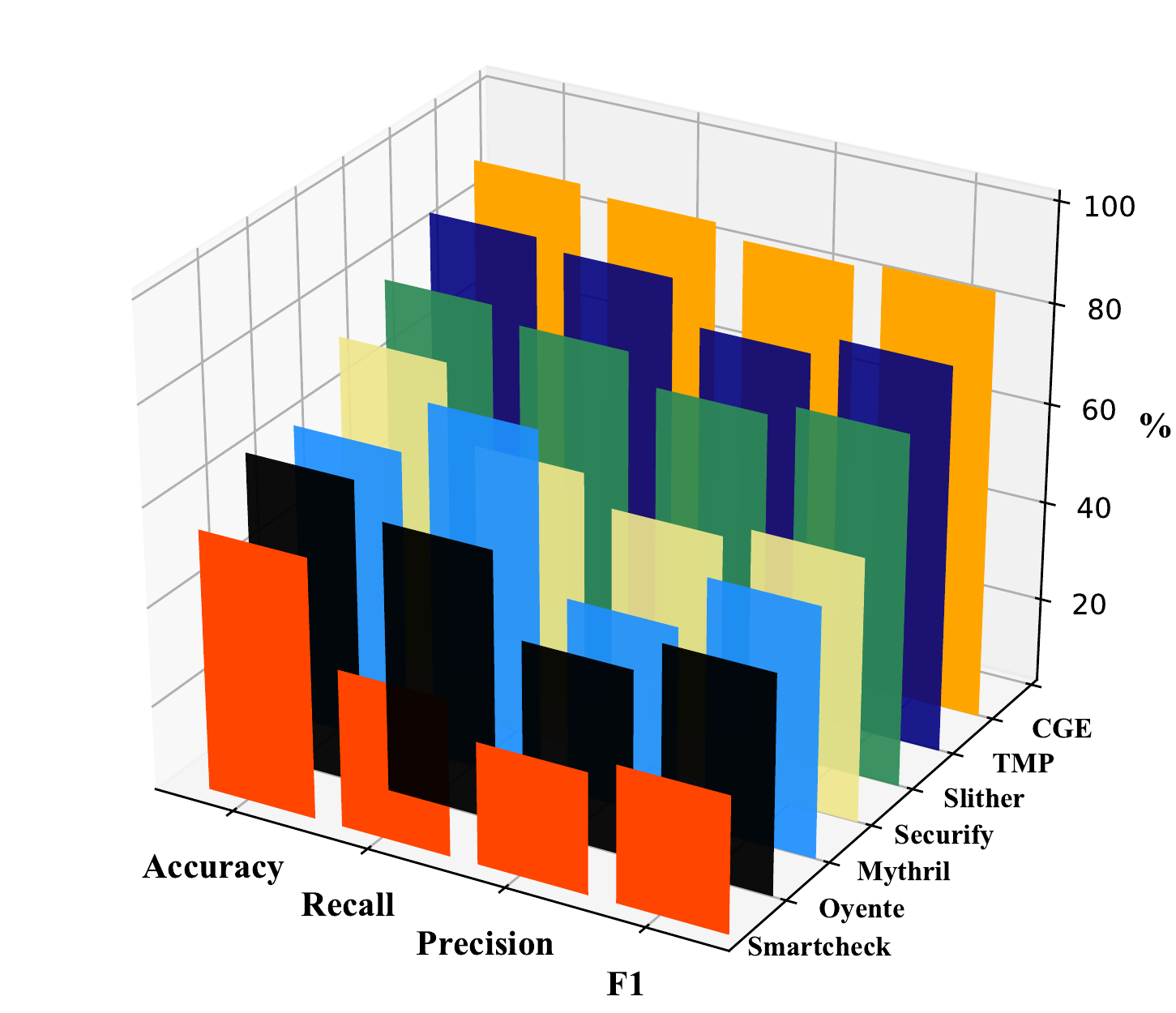}
}}
\resizebox{0.205\textwidth}{!}{
\subfigure[Timestamp comparison of tools]{
    \includegraphics[width=4.5cm]{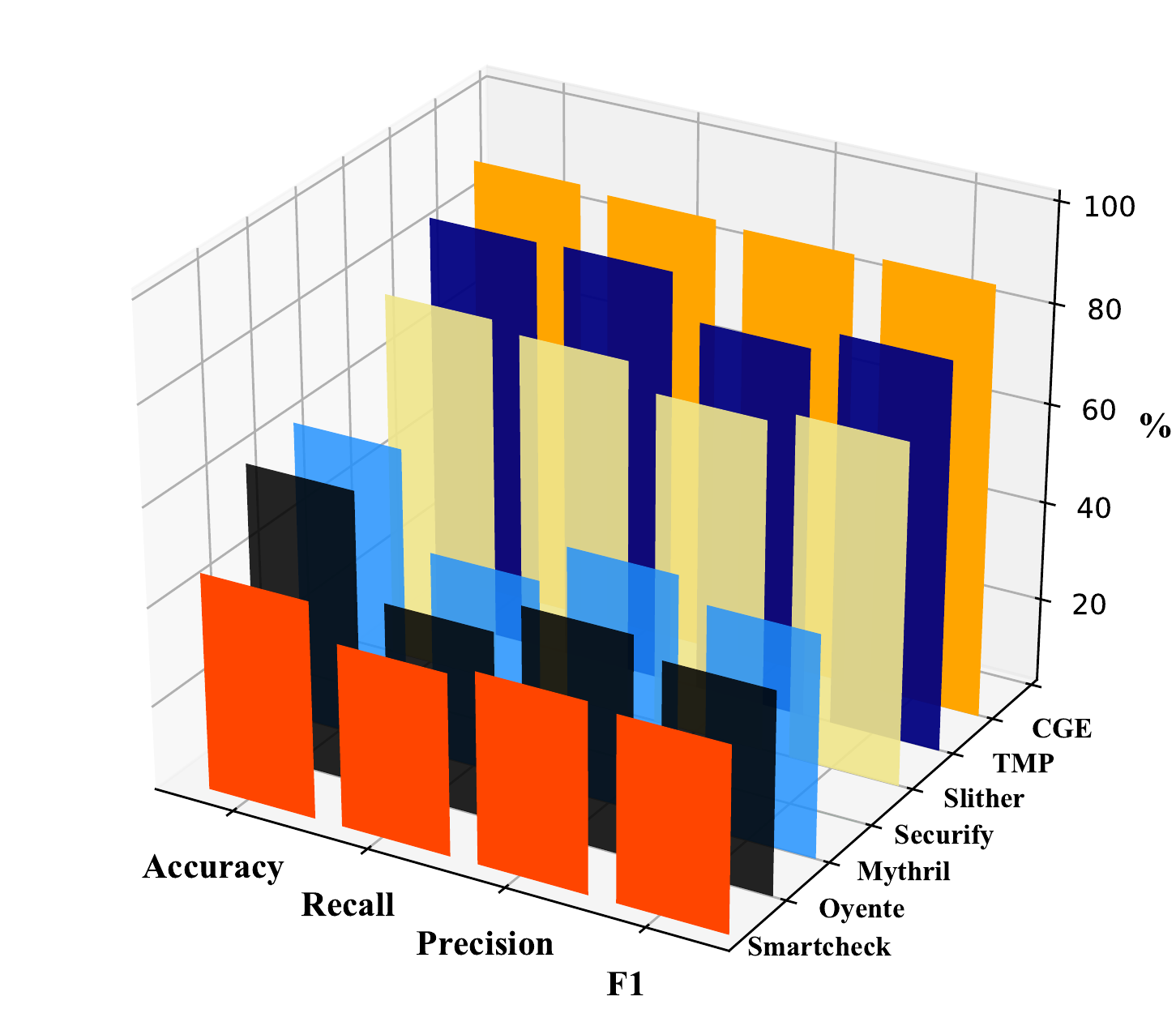}
}}
\resizebox{0.205\textwidth}{!}{
\subfigure[Infinite loop comparison of tools]{
    \includegraphics[width=4.5cm]{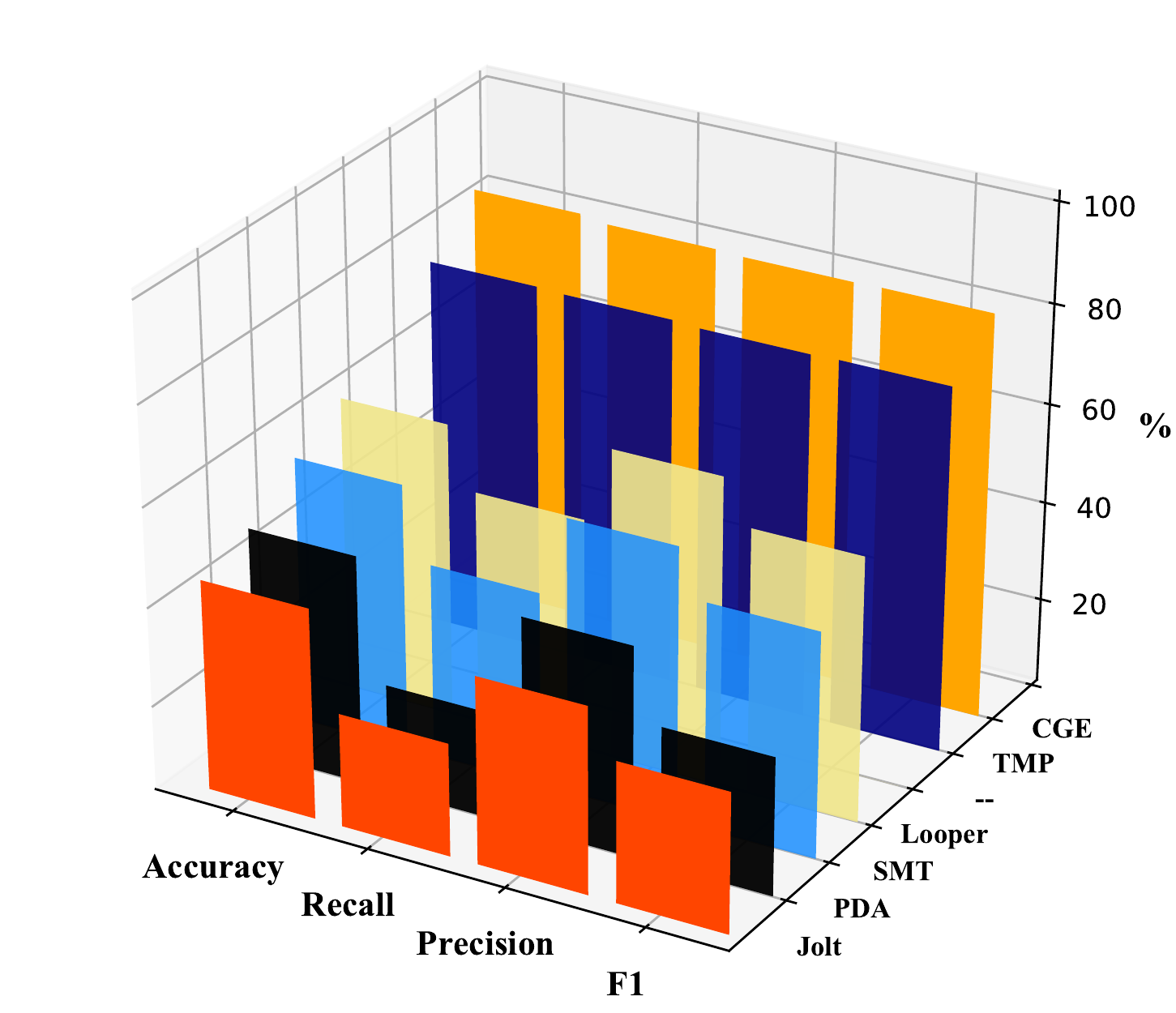}
}}
\quad\quad\quad\quad\quad\quad
\quad\quad\quad\quad\quad\quad
\resizebox{0.205\textwidth}{!}{
\subfigure[Reentrancy comparison of networks]{
    \includegraphics[width=4.5cm]{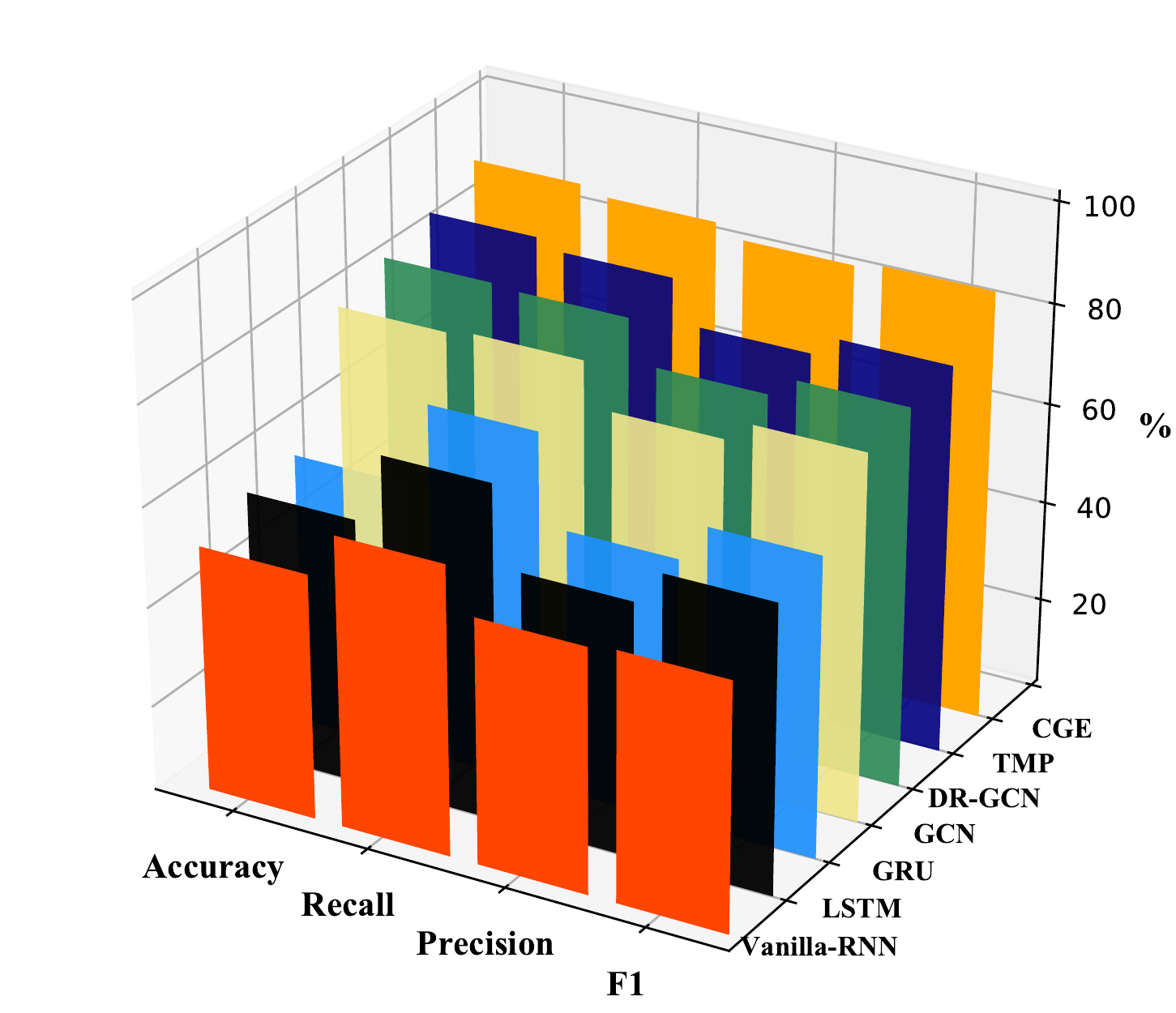}
}}
\resizebox{0.205\textwidth}{!}{
\subfigure[Timestamp comparison of networks]{
    \includegraphics[width=4.5cm]{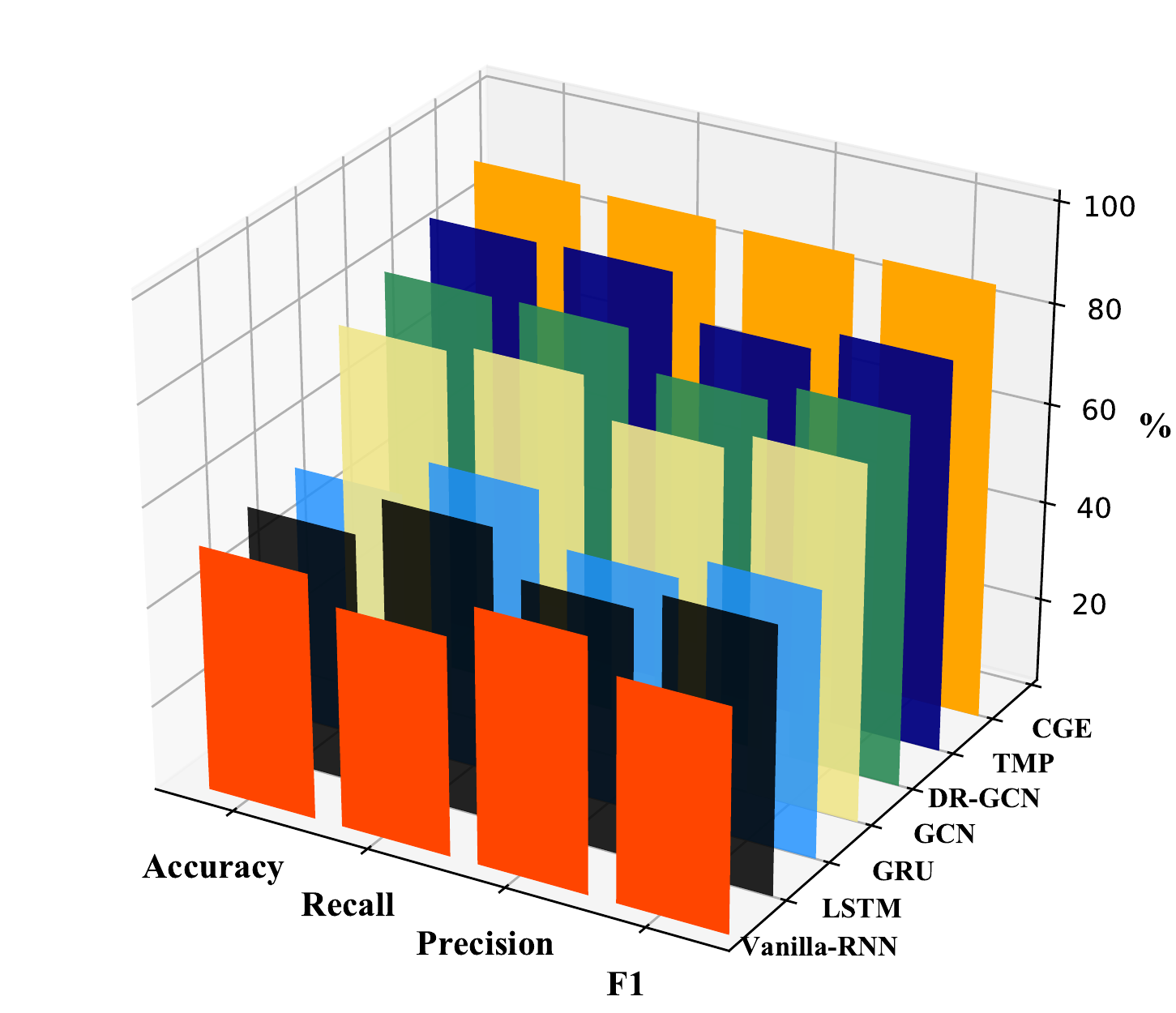}
}}
\resizebox{0.205\textwidth}{!}{
\subfigure[Infinite loop comparison of networks]{
    \includegraphics[width=4.5cm]{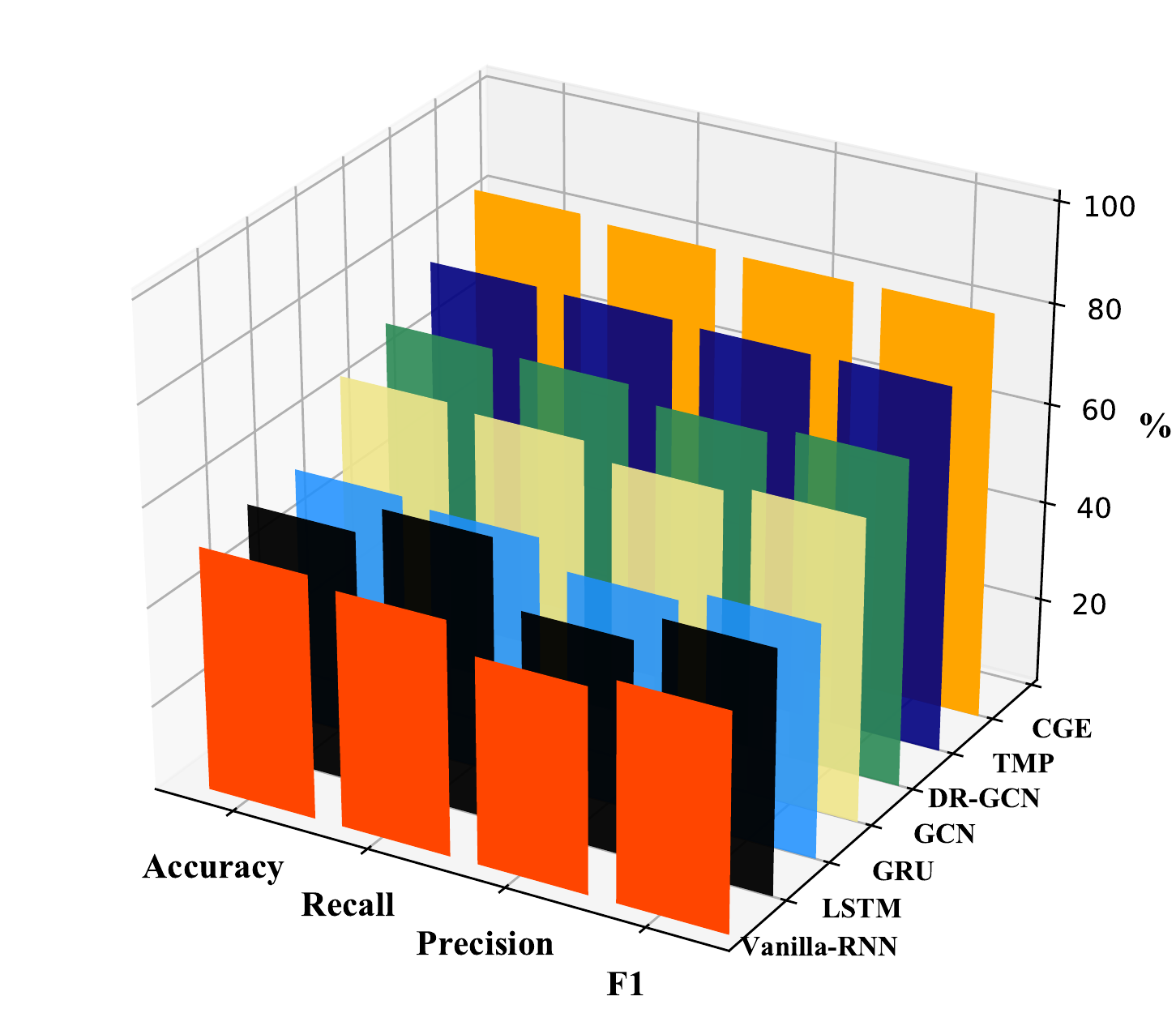}
}}
\caption{{Visuallization of the quantitative results in Table~\ref{Performance_comparison}: (a) \& (d) present comparison results of reentrancy vulnerability detection, while (b) \& (e) present comparison results of timestamp dependence detection, (c) \& (f) show comparison results of infinite loop vulnerability detection. In (a) \& (b), the 7 rows from front to back denote the Smartcheck, Oyente, Mythril, Securify, Slither, TMP, and CGE methods, respectively. In (c), the 6 rows from front to back denote the Jolt, PDA, SMT, Looper, TMP, and CGE methods, respectively. In (d) \& (e) \& (f), the 7 rows from front to back denote the Vanilla-RNN, LSTM, GRU, GCN, DR-GCN, TMP, and CGE methods, respectively. For each row in the figures, \emph{accuracy}, \emph{recall}, \emph{precision}, and \emph{F1 score} are respectively demonstrated from left to right.}}
\label{3D_fig}
 \vspace{-0.8em}
\end{figure*}

\textbf{Comparison on infinite loop vulnerability detection.} We also evaluated our methods on the infinite loop vulnerability. Specifically, we compare our methods against available infinite loop detection methods including:
\begin{itemize}[noitemsep,wide=0pt, leftmargin=\dimexpr\labelwidth + 2\labelsep\relax]
	\item \textbf{Jolt} \cite{Jolt}: The tool detects infinite loop bugs by monitoring the program state of two consecutive loop iterations. 
	\item \textbf{SMT} \cite{Smt}: An algorithm that relies on satisfiability modulo theories for automated detection of infinite loop bugs.
	\item \textbf{PDA} \cite{Pda}: A method that performs program path-based checking for infinite loop detection.
	\item \textbf{Looper} \cite{Looper}: Loop detection based on symbolic execution.
\end{itemize}

Quantitative results are illustrated in the right of Table~\ref{Performance_comparison}. From the table, we see that CGE consistently and significantly outperforms other methods on the infinite loop vulnerability detection task. In particular, CGE achieves a 83.21\% accuracy and a 82.13\% F1 score. In contrast, state-of-the-art detection tools Looper are 59.56\% and 53.87\%, and TMP are 74.61\% and 74.10\%. The improvements may come from the fact that we consider key variables and rich dependencies between program elements in smart contracts. 

We further visualize the quantitative results of Table~\ref{Performance_comparison} in Figs.~\ref{3D_fig}(a), (b), and (c). Specifically, Fig.~\ref{3D_fig}(a) and Fig.~\ref{3D_fig}(b) present comparison results of reentrancy vulnerability detection and timestamp dependence vulnerability detection, respectively. The 7 rows (in different colors) from front to back denote methods \textit{Smartcheck}, \textit{Oyente}, \textit{Mythril}, \textit{Securify}, \textit{Slither}, TMP, and CGE, respectively. For each row in the figures, \emph{accuracy}, \emph{recall}, \emph{precision}, and \emph{F1 score} are respectively demonstrated from left to right. Fig.~\ref{3D_fig}(c) shows comparison results of infinite loop vulnerability detection, where the 6 rows from front to back denote \textit{Jolt}, \textit{PDA}, \textit{SMT}, \textit{Looper}, TMP, and CGE methods, respectively. We can clearly observe that CGE outperforms existing methods by a large margin.

\vspace{-0.7em}
\subsection{A Case Study Towards Better Understanding of the Reasons Behind the Results (RQ2)}
In this subsection, we present an interesting case of smart contract vulnerabilities, which may bring new insights into the abilities of the studied methods. Particularly, we investigate a new type of reentrancy vulnerability, \emph{i.e., sharing-variable} reentrancy. To our knowledge, most existing methods cannot precisely detect such vulnerabilities.

Besides classical reentrancy introduced in Fig.~\ref{fig_dao} and section~\ref{sec_problem}, a reentrancy attack is also possible when a transfer function shares internal variables with another function, which we define as sharing-variable reentrancy.

\begin{figure}
 	\centering
 	\includegraphics[width=3in]{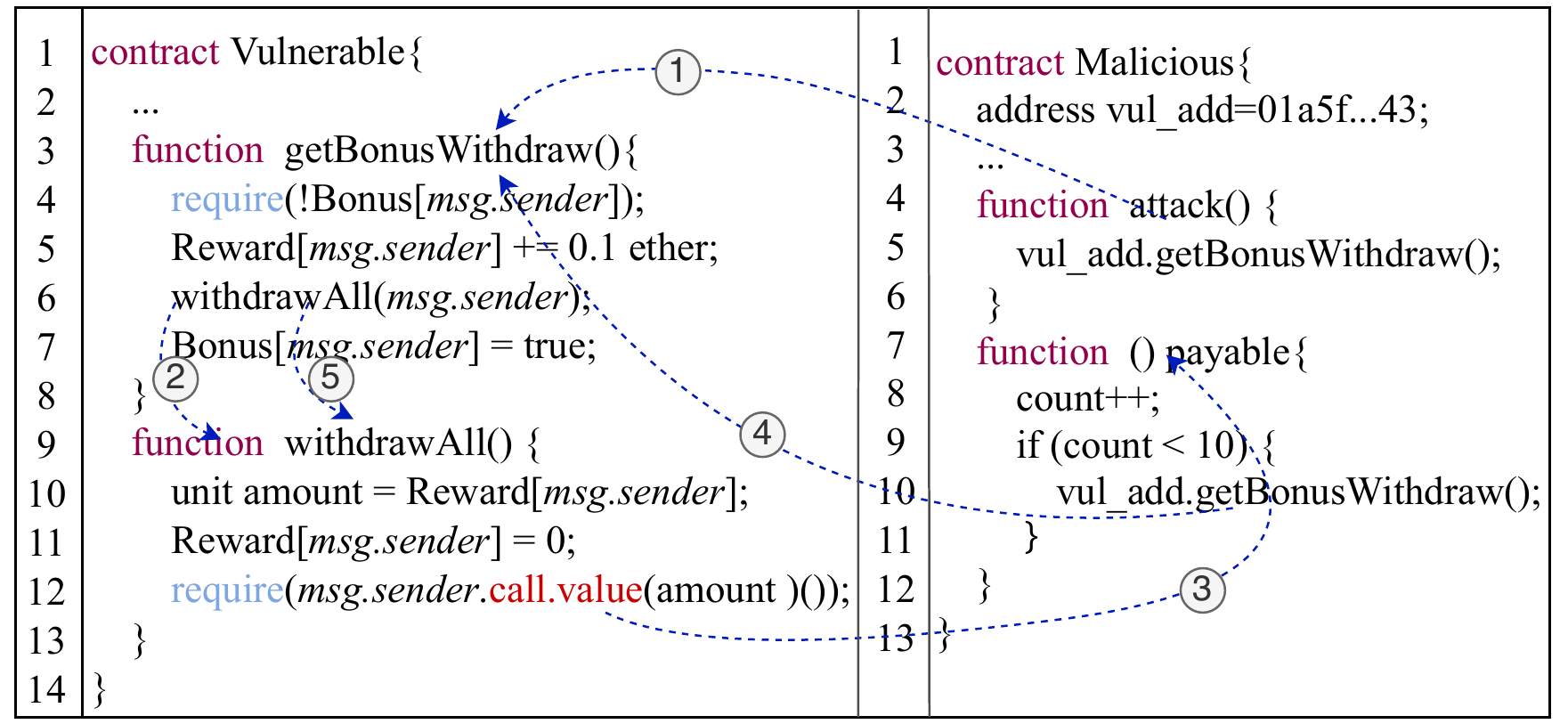}
 	\caption{A real-world smart contract with the sharing-variable reentrancy vulnerability.}
 	\label{fig:sharingvariable}
	 \vspace{-0.8em}
\end{figure}

In Fig. \ref{fig:sharingvariable}, we illustrate a real-world sharing-variable reentrancy example, where the \emph{Malicious} contract plays an attack role against the \emph{Vulnerable} contract. More specifically, contract \emph{Vulnerable} contains two functions: \emph{getBonusWithdraw} and \emph{withdrawAll}. Function \emph{withdrawAll} allows a user to withdraw all her rewards, while function \emph{getBonusWithdraw} allows a user to withdraw all her rewards together with a 0.1 Ether bonus for each new user. 

\textbf{Attack.} As demonstrated in Fig. \ref{fig:sharingvariable}, contract \emph{Malicious} first uses its \emph{attack} function to call the $\emph{getBonusWithdraw}$ function of contract \emph{Vulnerable} (step 1). As $\emph{getBonusWithdraw}$ invokes the \emph{withdrawAll} function (\emph{Vulnerable}, line 6) to send the rewards and bonus to \emph{Malicious} (step 2). This will automatically trigger the fallback function of \emph{Malicious} (step 3), where \emph{Malicious} invokes $\emph{getBonusWithdraw}$ again to steal money (step 4). Since the bonus flag \emph{Bonus}[\emph{msg.sender}] has yet been set to true, \emph{Vulnerable} believes \emph{Malicious} has not got the new user bonus yet and thus gives 0.1 Ether bonus again to \emph{Vulnerable} (\emph{Vulnerable}, line 5), then function \emph{withdrawAll} is re-entered to withdraw the 0.1 Ether illegal bonus (step 5). \emph{Malicious} actually invokes \emph{getBonusWithdraw} 9 times (\emph{Malicious}, line 9) in its fallback function to steal 1 Ether.

\textbf{Underlying issue.} This example reveals that although in the \emph{withdrawAll} function, contract \emph{Vulnerable} updates the user balance (i.e., \emph{Reward}) before money transfer, \emph{Malicious} can still be attacked. The novel attack utilizes the shared variable (\emph{Reward}) to steal money. Although \emph{withdrawAll} function itself is safe, the malicious contract may call $\emph{getBonusWithdraw}$ to modify the shared variable \emph{Reward} to enable attacks.

Unfortunately, such kind of attacks cannot yet be detected by existing methods. We empirically checked the \emph{Vulnerable} contract using the state-of-the-art tools including \emph{Oyente \cite{oyente}, Securify \cite{securify}, {Smartcheck} \cite{smartcheck}, Slither \cite{feist2019slither}, and Mythril \cite{mythril}}, and manually inspected their generated reports. Oyente, Smartcheck, Slither, and Mythril fail to identify the reentrancy bug, whereas Security presents a lot of warnings all at the wrong places and misses the sharing-variable reentrancy vulnerability as well. In contrast, CGE successfully detects the vulnerability. These evidences reveal that the underlying detection rules of existing reentrancy vulnerability detection methods indeed can be cheated by the \emph{sharing variable} trick and some vulnerability patterns are hard to be covered. The current rules check only the \emph{user balance} variable that is directly related to the \emph{call.value} invocation, while ignoring dependencies between variables, e.g., other variables may affect the \emph{user balance} variable. In this regard, an essential highlight of our method is the capability of capturing data dependencies between critical variables.

\vspace{-0.7em}
\subsection{Comparison with Neural Network-based Methods (RQ3)} 
We further compare our methods with other neural network alternatives to seek out which neural network architectures could succeed in the smart contract vulnerability detection task. The compared methods are summarized below.
\begin{itemize}[noitemsep,wide=0pt, leftmargin=\dimexpr\labelwidth + 2\labelsep\relax]
 \item \textbf{Vanilla-RNN} \cite{Vanilla-RNN}: A two-layer recurrent neural network, which takes the code sequence as input and evolves its hidden states recurrently to capture the sequential pattern lying in the code.
 \item \textbf{LSTM} \cite{LSTM}: The most widely used recurrent neural network for processing sequential data. LSTM is short for long short term memory, which recurrently updates the cell state upon successively reading the code sequence.
 \item \textbf{GRU} \cite{GRU}: The gated recurrent unit, which uses gating mechanisms to handle the code sequence.
 \item \textbf{GCN} \cite{GCN}: Graph convolutional network that takes the contract graph as input and implements layer-wise convolution on the graph using  graph Laplacian.
\item \textbf{DR-GCN} \cite{ijcai}: The degree-free graph convolutional network, which increases the connectivity of nodes and removes the diagonal node degree matrix.
 \item \textbf{TMP} \cite{ijcai}: The temporal message propagation network, which learns the contract graph feature by flowing information along the edges successively following their temporal order. The final graph feature is used for vulnerability prediction.
\end{itemize}

For a feasible comparison, Vanilla-RNN, LSTM, and GRU are fed with the contract function code sequence, represented as vectors. GCN, DR-GCN, and TMP are presented with the normalized graph extracted from the source code and are required to detect the corresponding vulnerabilities. 

We illustrate the results of different models in terms of \emph{accuracy}, \emph{recall}, \emph{precision}, and \textit{F1 score} in Table~\ref{Performance_comparison}, while Figs.~\ref{3D_fig}(d), (e), and (f) further visualize the results. Interestingly, experimental results show that Vanilla-RNN, LSTM, and GRU perform relatively worse than the state-of-the-art conventional (non-deep-learning) methods. In contrast, graph neural networks GCN, DR-GCN, and TMP, which are capable of handling graphs, achieve significantly better results than conventional methods. This suggests that blindly treat the source code as a sequence is not suitable for the vulnerability detection task, while modeling the source code into graphs and adopting graph neural networks is promising. We conjecture that processing code sequentially loses valuable information from smart contract code since they ignore the structural information of contract programs, such as the data-flow and invocation relationships. The accuracies of GCN and DR-GCN are lower than TMP,  this may due to the fact that GCN and DR-GCN fail to capture the temporal information induced by data flow and control flow, which is explicitly considered in TMP using ordered edges. Further, we attribute the improved performance of CGE over TMP to that TMP does not consider known security patterns and ignores key variables.

\renewcommand\arraystretch{1.0}
\begin{table*}
\centering
\resizebox{0.95\textwidth}{!}{
\begin{tabular}{lcccccccccccc}
\toprule
{\multirow{3}{*}{\textbf{Metrics}}} & \multicolumn{4}{c}{\textbf{Reentrancy}} & \multicolumn{4}{c}{\textbf{Timestamp dependence}} & \multicolumn{4}{c}{\textbf{Infinite loop}} \\
\cmidrule(lr){2-5}\cmidrule(lr){6-9}\cmidrule(lr){10-13} & CGE-WOG &  CGE-WOE & CGE-WON & CGE & CGE-WOG & CGE-WOE &CGE-WON & CGE & CGE-WOG & CGE-WOE &CGE-WON & CGE  \\
\midrule
Acc(\%) & 82.09 & 84.42 & 86.34 & \textbf{89.15} & 81.30  & 83.52 &  86.61 & \textbf{89.02} & 72.23 & 74.68 & 79.51 & \textbf{83.21} \\
Recall(\%) &  80.18 &  82.65 & 84.38 & \textbf{87.62} & 80.68  &  82.89 & 84.06 & \textbf{88.10} & 70.08  & 74.21  & 77.14 & \textbf{82.29} \\
Precision(\%) & 72.15 & 78.94 & 83.35 & \textbf{85.24} & 78.42  & 80.16  & 83.90 & \textbf{87.41} & 71.44 & 73.86  & 76.26 & \textbf{81.97} \\
F1(\%) & 75.95  &  80.75 & 83.86 & \textbf{86.41} & 79.53 & 81.50  & 83.98 & \textbf{87.75} & 70.75  & 74.03 & 76.70 & \textbf{82.13} \\
\bottomrule
\end{tabular}
}
\caption{Accuracy comparison between CGE and its variants on the three vulnerability detection tasks.}
\label{normalization_com}
\end{table*}

\renewcommand\arraystretch{1.0}
\begin{table*}
\centering
\resizebox{0.95\textwidth}{!}{
\begin{tabular}{lcccccccccccc}
\toprule
\multirow{2}{*}{\textbf{Variants}} & \multicolumn{4}{c}{\textbf{Reentrancy}} & \multicolumn{4}{c}{\textbf{Timestamp dependence}} &  \multicolumn{4}{c}{\textbf{Infinite Loop}}\\
\cmidrule(lr){2-5}\cmidrule(lr){6-9}\cmidrule(lr){10-13} & Acc(\%) & Recall(\%) & Precision(\%) & F1(\%) & Acc(\%) & Recall(\%) & Precision(\%) & F1(\%) & Acc(\%) & Recall(\%) & Precision(\%) & F1(\%) \\
\midrule
Vanilla-RNN-EP & 56.06 & 60.24 & 58.21 & 59.20 & 54.58 & 49.65 & 59.35 & 54.07 & 54.72 & 52.62 & 49.94 & 51.24 \\
LSTM-EP & 60.15 & 72.26 & 58.68 & 64.77 & 59.82 & 63.38 & 58.28 & 56.29 & 56.52 & 59.98 & 49.75 & 54.39 \\
GRU-EP & 62.08 & 75.01 & 60.13 & 66.75 & 61.22 & 64.18 & 58.45 & 61.18 & 57.09 & 60.54 & 49.81 & 54.65 \\
GCN-EP & 80.96 & 81.05 & 76.84 & 78.89 & 79.32 & 79.94 & 73.65 & 76.67 & 70.06 & 69.81 & 64.29 & 66.94 \\	
DR-GCN-EP & 85.14 & 84.12 & 79.38 & 81.68 & 83.74 & 84.02 & 80.59 & 82.27 & 74.36 & 73.08 & 69.45 & 71.22 \\
\midrule
CGE(LSTM) & 86.74 & 85.18 & 82.85 & 84.00 & 87.92 & 85.08 & 87.13 & 86.09 & 79.18 & 78.25 & 76.80 & 77.52\\
CGE(FC) & 87.64 & 85.74 & 82.97 & 84.33 & 88.12 & 87.98 & 85.04 & 86.49 & 80.62 & 78.96 & 77.24 & 78.09 \\
CGE(1-FC) & 88.54 & 86.12 & 83.80 & 84.94 & 86.62 & 87.82 & 81.73 & 84.66 & 81.43 & 81.25 & 80.98 & 81.11 \\
CGE(2-FC) & 88.89 & 86.47 & 84.51 & 85.48 & 87.05 & 84.96 & 85.02 & 84.98 & 81.82 & 81.76 & 80.54 & 81.15 \\
CGE(AP) & 88.02 & 85.92 & 83.45 & 84.67 & 85.25 & 85.16 & 81.84 & 83.47 & 79.53 & 78.58 & 76.94 & 77.75 \\	
\textbf{CGE} & \textbf{89.15} & \textbf{87.62} & \textbf{85.24} & \textbf{86.41} & \textbf{89.02} &  \textbf{88.10} & \textbf{87.41} & \textbf{87.75} & \textbf{83.21} & \textbf{82.29} & \textbf{81.97} & \textbf{82.13} \\
\bottomrule
\end{tabular}
}
\caption{Upper: Performance comparison between CGE and other neural networks combined with expert patterns. ‘-EP’ denotes combining with expert patterns. Lower: Comparison with other feature fusion network architectures.}
\label{expert_pattern_com}
 \vspace{-1.5em}
\end{table*}

\begin{figure*}
\centering 
\resizebox{0.95\textwidth}{!}{
\subfigure[Reentrancy]{
    \begin{minipage}[t]{0.342\linewidth}
    \includegraphics[width=7.0cm]{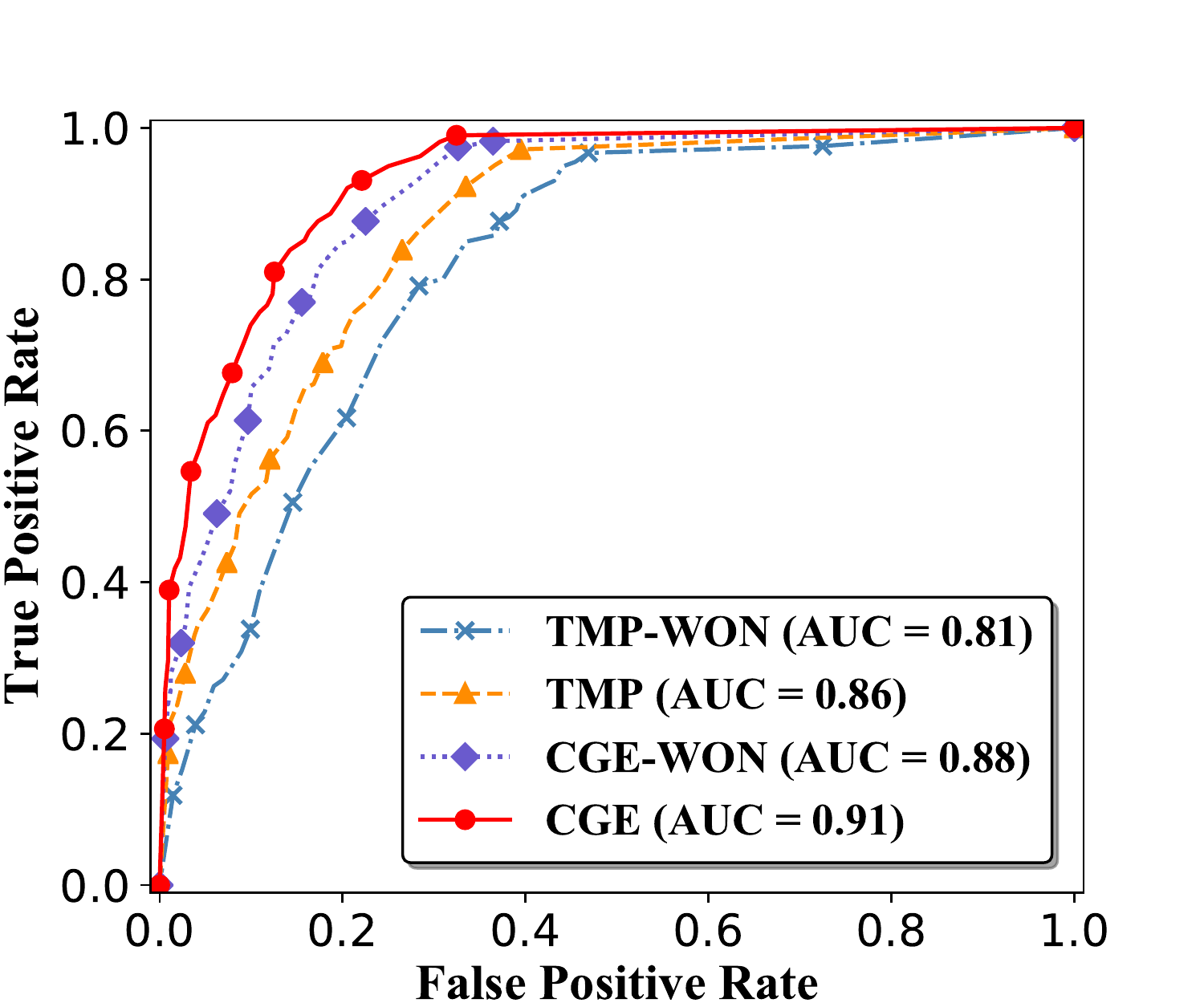}
    \end{minipage}
}
\quad
\subfigure[Timestamp dependence]{
    \begin{minipage}[t]{0.35\linewidth}
    \includegraphics[width=7.0cm]{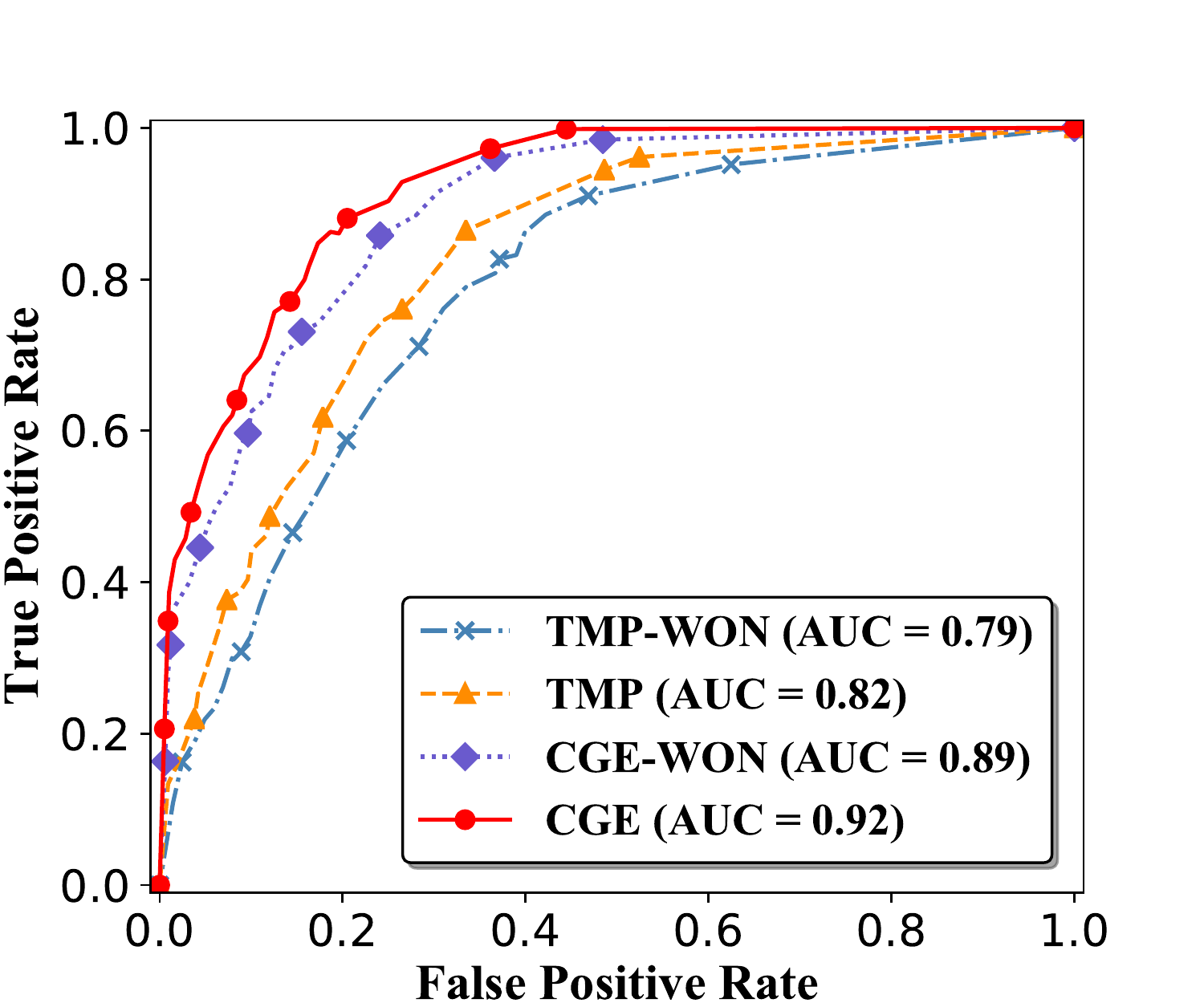}
    \end{minipage}
}
\quad
\subfigure[Infinite loop]{
    \begin{minipage}[t]{0.35\linewidth}
    \includegraphics[width=7.0cm]{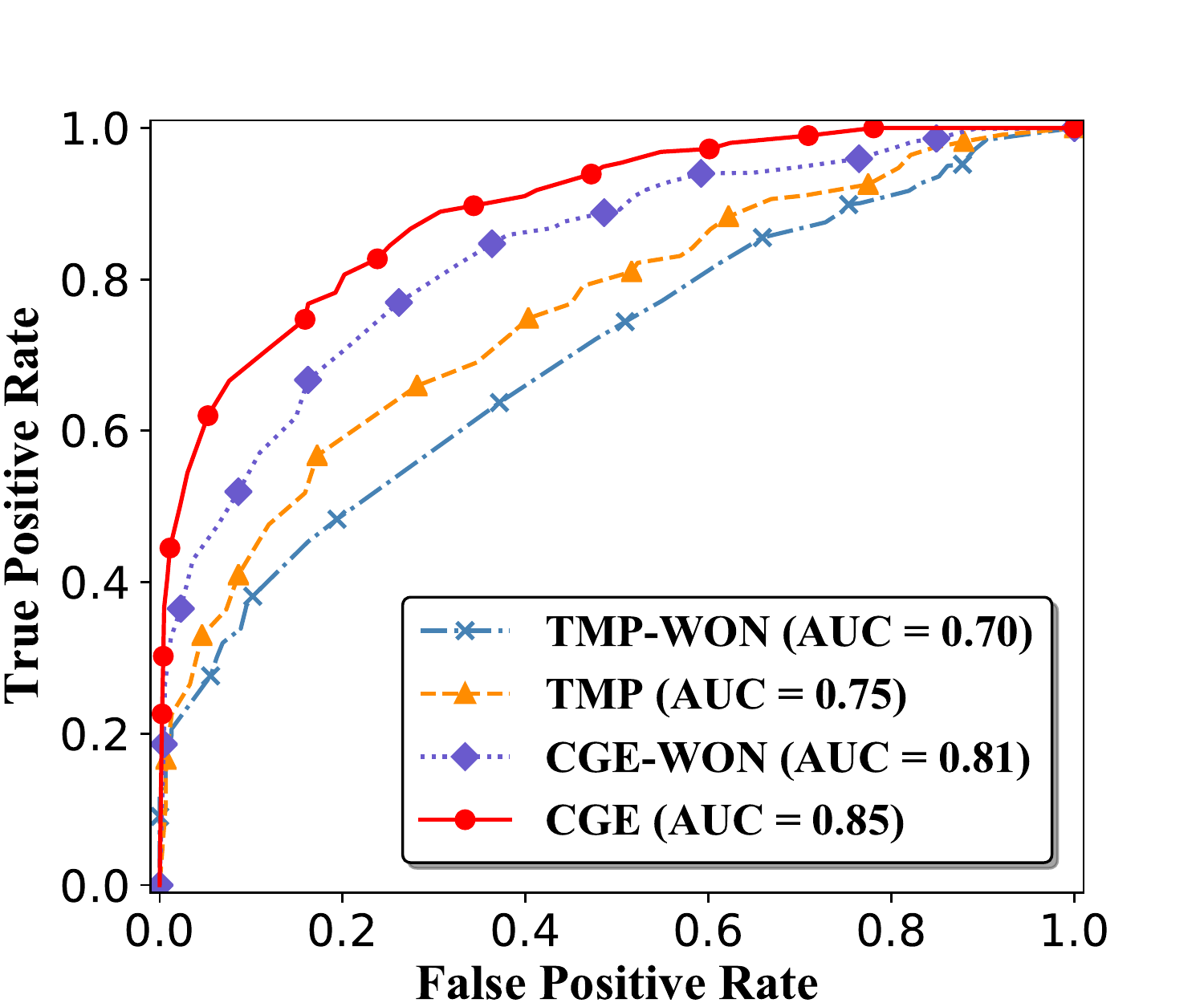}
    \end{minipage}
}
\quad
\subfigure[Accuracy study on the security pattern module]{
    \begin{minipage}[t]{0.35\linewidth}
    \includegraphics[width=7.0cm]{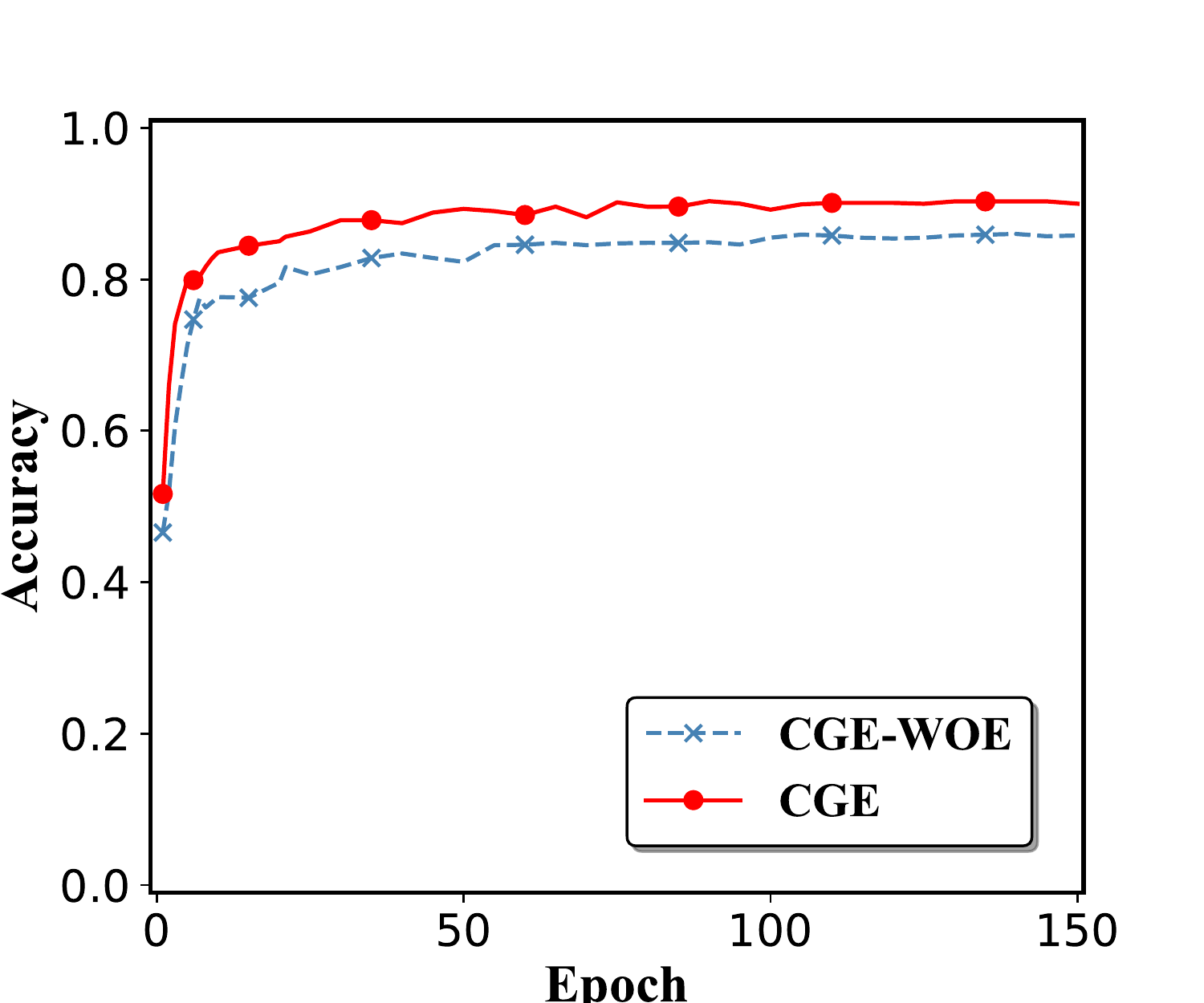}
    \end{minipage}
}
\quad
\subfigure[Accuracy study on the graph feature module]{
    \begin{minipage}[t]{0.35\linewidth}
    \includegraphics[width=7.0cm]{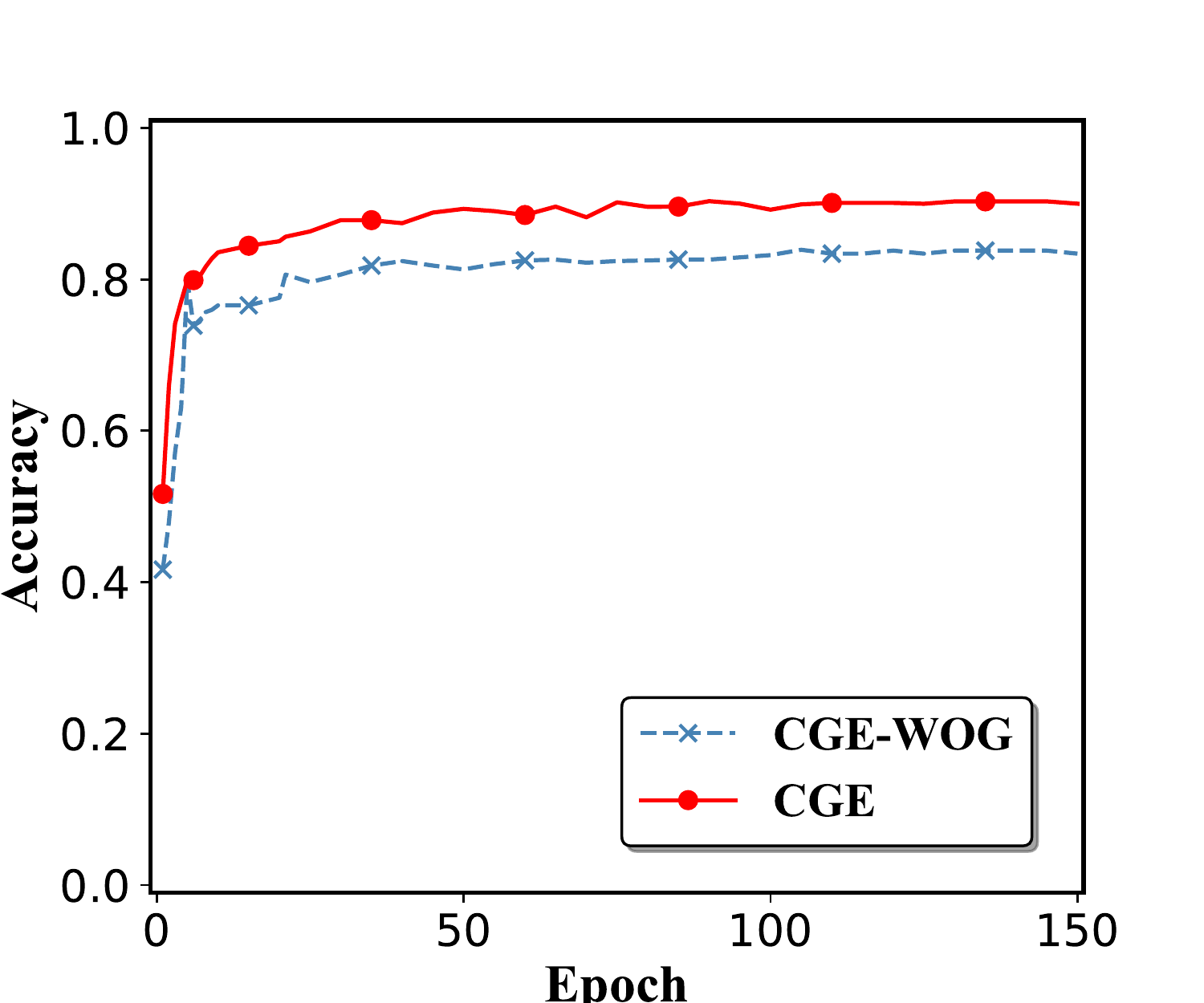}
    \end{minipage}
}
}
\caption{Curves comparison: (a), (b), and (c) present the ROC analysis of graph normalization module for TMP, CGE, and their variants on the three vulnerability detection tasks, where AUC stands for area under the curve. In (d), the two curves study the effect of removing the security pattern extraction module, while (e) presents the study on removing the contract-graph feature extraction module.}
\label{fig_roc}
\vspace{-1.8em}
\end{figure*}

\vspace{-0.7em}
\subsection{Ablation Study (RQ4)}
By default, CGE adopts the \emph{graph normalization} module to highlight the core nodes in the \emph{contract graph},  it is interesting to study the effect of removing this module. Moreover, CGE incorporates an \emph{expert pattern extraction} module and a \emph{message propagation} module to aggregate information from both security patterns and the contract graph. It is useful to evaluate the contributions of the two modules by removing them respectively from CGE. Finally, we are also interested in exploring the effect of different network layers in CGE. In what follows, we conduct experiments to study the four aforementioned modules.

\textbf{Effect of the graph normalization module.} We removed the graph normalization module (introduced in subsection~\ref{graph_normalization}) from CGE, and compared it with the default CGE. The variant is denoted as CGE-WON, where WON is short for \emph{without normalization}. Quantitative results are summarized in Table~\ref{normalization_com}. We can observe that with the proposed graph normalization phase, the performance of CGE is better. For example, for reentrancy vulnerability detection task, the CGE model obtains a 2.81\% and 2.55\% improvement in terms of accuracy and F1 score, respectively.

Figs.~\ref{fig_roc}(a) \& (b) \& (c) further plot the ROC curves of CGE and CGE-WON. We adopt Receiver Operating Characteristic (ROC) analysis to show the impact of the graph normalization module. AUC (area under the curve) is used as the measure for performance, the higher AUC the better performance. Fig.~\ref{fig_roc}(a) demonstrates that CGE performs better on the reentrancy detection task, the AUC increases by 0.03 with the graph normalization module. On the timestamp dependence detection task, CGE obtains a 0.03 improvement in AUC (shown in Fig. ~\ref{fig_roc}(b)). On the infinite loop detection task, CGE gains a 0.04 improvement in AUC (shown in Fig.~\ref{fig_roc}(c)). In the figures, we also demonstrate the effect of removing the graph normalization module of another method, namely TMP. Similar findings are observed. The experimental results suggest that program elements should contribute distinctly to vulnerability detection rather than having equal contributions.

\textbf{Effect of the security pattern module.} To evaluate the effect of our proposed security pattern module, we analyze the performance of CGE with and without the security pattern module. Towards this, we modify CGE by removing the expert pattern extraction module, utilizing only the graph feature for vulnerability learning and detection. This variant is denoted as CGE-WOE, where WOE is short for without expert pattern. The empirical findings are demonstrated in Table~\ref{normalization_com}, while the visual curves are illustrated in Fig. \ref{fig_roc}(d). In Fig. \ref{fig_roc}(d), the red curve demonstrates the accuracy of CGE over different epochs on the reentrancy vulnerability detection. Obviously, we can observe that the performance of CGE is consistently superior to CGE-WOE across all epochs, revealing that incorporating security patterns is necessary and important to improve the performance. Quantitative results on all the three vulnerabilities, which are presented in Table~\ref{normalization_com}, further reconfirm the finding. 

We also conduct experiments to extend other neural networks with expert patterns, and empirically compared these methods with CGE. The results are illustrated in Table~\ref{expert_pattern_com}, where `-EP' denotes combining with expert patterns. We can observe that neural networks combined with expert patterns indeed achieve better results compared to their pure neural network counterparts. For example, DR-GCN-EP gains a 4.92\% accuracy improvement over DR-GCN in average, and LSTM-EP obtains a 6.91\% accuracy improvement over LSTM. These results indicate the effectiveness of combining neural networks with expert patterns. On the other hand, the proposed method CGE consistently outperforms other approaches including DR-GCN-EP. DR-GCN-EP ranks second in the tested methods.

\textbf{Effect of the contract graph feature extraction module.} We further investigate the impact of the contract graph feature extraction module in CGE by comparing it with its variant. Towards this, we remove the proposed contract graph construction and temporal message propagation module, while utilizing only the security pattern feature. The new variant is denoted as CGE-WOG, namely CGE without contract graph feature. Fig. \ref{fig_roc}(e) visualizes the results, where the red curve demonstrates the accuracy of CGE over different epochs, while the blue curve shows the accuracy of CGE-WOG. Clearly, the performance of CGE is consistently better compared to its variant across all epochs. Quantitative results are further presented in Table~\ref{normalization_com}, where all the three vulnerabilities are involved. The results, together with the experimental results on CGE-WOE, suggest that the contract graph feature contributes significant performance gain in CGE and leads to a higher gain than the security pattern feature. 

 \textbf{Effect of different feature fusion networks.} When combining security pattern features and contract graph features, CGE uses a neural network with a convolution layer and a max pooling layer followed by 3 fully connected layers and a sigmoid layer. To verify this network architecture, we also study five other alternatives. First, we replace the convolution and max pooling layer with a fully connected layer, which we denote as CGE(FC). We also try replacing them with an LSTM layer, which we term as CGE(LSTM). Then, we keep the convolution and max pooling layer, but change the 3 fully connected layers to 1 or 2 fully connected layers. The two variants are denoted as CGE(1-FC) and CGE(2-FC), respectively. Finally, we explore replacing the max pooling layer with an average pooling layer, namely CGE(AP), while keeping other layers fixed. The empirical results are illustrated in Table~\ref{expert_pattern_com}. The  results reveal that: 1) RNN architectures such as LSTM are not suitable for the feature fusion task, 2) the default setting of CGE yields better results than the five alternatives, and 3) using average pooling or changing the number of fully connected layers leads to a slight performance drop.

\vspace{-0.7em}
\section{Discussions}
\textbf{Specialty of our method in dealing with smart contracts.} Distinct from conventional programs that consume only CPU resources, users have to pay a fee for executing each line of smart contract code. The fee is approximately proportional to how much code needs to run and is referred to as \emph{gas}. Therefore, in the proposed method, we studied the infinite loop vulnerability since an \emph{infinite loop} will consume a lot of gas but all the gas is consumed in vain. This is because the infinite loop is unable to change any state (any execution that runs out of gas is aborted).  Moreover, the function libraries of the smart contracts and other program languages are quite different. For example, \emph{call.value} and \emph{block.timestamp} are unique and specially designed in smart contracts. We implement an open-sourced tool to analyze the specific syntax of smart contract statements. We also use core nodes to symbolize  invocations and variables closely related to a specific vulnerability, and represent other variables and invocations as normal nodes. We would like to point out that there is a unique fallback mechanism in smart contracts, which is different from other programming languages. In the contract graph, we build a fallback node to stimulate the fallback function of a virtual attack contract, which can interact with the function under test.

\textbf{Discussions on the contract graph.} Existing efforts adopted the \emph{control flow graph}, \emph{code property graph}, and \emph{abstract syntax tree} to represent program code. The differences between them and our \emph{contract graph} can be summarized as: (i) Control flow graph utilizes a node to model a basic block, \emph{i.e. a straight-line piece of code without any jumps,} and {uses} edges to represent jumps \cite{phan2017convolutional}. They focus mainly on execution path jumps and tend to consider each node as of equal importance. (ii) \emph{Code property graph} \cite{yamaguchi2014modeling,suneja2020learning} models statements as nodes, and represents the control flow between statements as edges. (iii) \emph{Abstract syntax tree} \cite{mou2016convolutional,zhang2019novel} adopts a tree representation of the abstract syntactic structure of source code, which relies on a tree structure and has difficulties in fully characterizing the rich semantic information between nodes. (iv) In our contract graph, nodes are used to model variables and invocations related to a specific vulnerability and are classified into different categories, i.e. \emph{core nodes, normal nodes}, and \emph{fallback nodes}. We also explicitly model the order of the edges following their temporal order in the code and consider the specific fallback mechanism of the smart contracts.

\vspace{-0.7em}
\section{Conclusion and Future Work}
\label{conclusion}
In this paper, we have proposed a fully automated approach for smart contract vulnerability detection at the function level. In contrast to existing approaches, we combine both expert patterns and contract graph semantics, consider rich dependencies between program elements, and explicitly model the fallback mechanism of smart contracts. We also explore the possibility of using novel graph neural networks to learn the graph feature from the contract graph, which contains rich control- and data- flow semantics. Extensive experiments are conducted, showing that our method significantly outperforms the state-of-the-art vulnerability detection tools and other neural network-based methods. We believe our work is an important step towards revealing the potential of combining deep learning with conventional patterns on smart contract vulnerability detection tasks. For future work, we will investigate the possibility of extending this method to smart contracts that have only bytecode, and explore this architecture on more other vulnerabilities.

\vspace{-0.7em}
\section*{Acknowledgments}
This paper was supported by the Natural Science Foundation of Zhejiang Province, China (Grant No. LQ19F020001), the National Natural Science Foundation of China (No. 61902348, 61802345), and the Research Program of Zhejiang Lab (2019KD0AC02).

\ifCLASSOPTIONcaptionsoff
  \newpage
\fi


%

\footnotesize
\bibliographystyle{IEEEtran}
\bibliography{bare_jrnl_compsoc}

%

\vspace{-4em}
\begin{IEEEbiography}[{\includegraphics[width=1in,height=1.25in,clip,keepaspectratio]{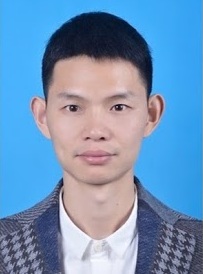}}]{Zhenguang Liu}
is currently a professor of Zhejiang Gongshang University. He had been a research
fellow in National University of Singapore and A*STAR. 
He respectively received his Ph.D. and B.E. degrees from Zhejiang University
and Shandong University, China. His research interests include smart contract security and multimedia
data analysis. 
Dr. Liu
has served as technical program committee member for conferences
such as ACM MM, CVPR, AAAI, IJCAI, and ICCV, session chair of ICGIP, local chair of KSEM, and reviewer for IEEE TVCG, IEEE TPDS, ACM TOMM, etc.  
\end{IEEEbiography}

\vspace{-4em}
\begin{IEEEbiography}[{\includegraphics[width=1in,height=1.25in,clip,keepaspectratio]{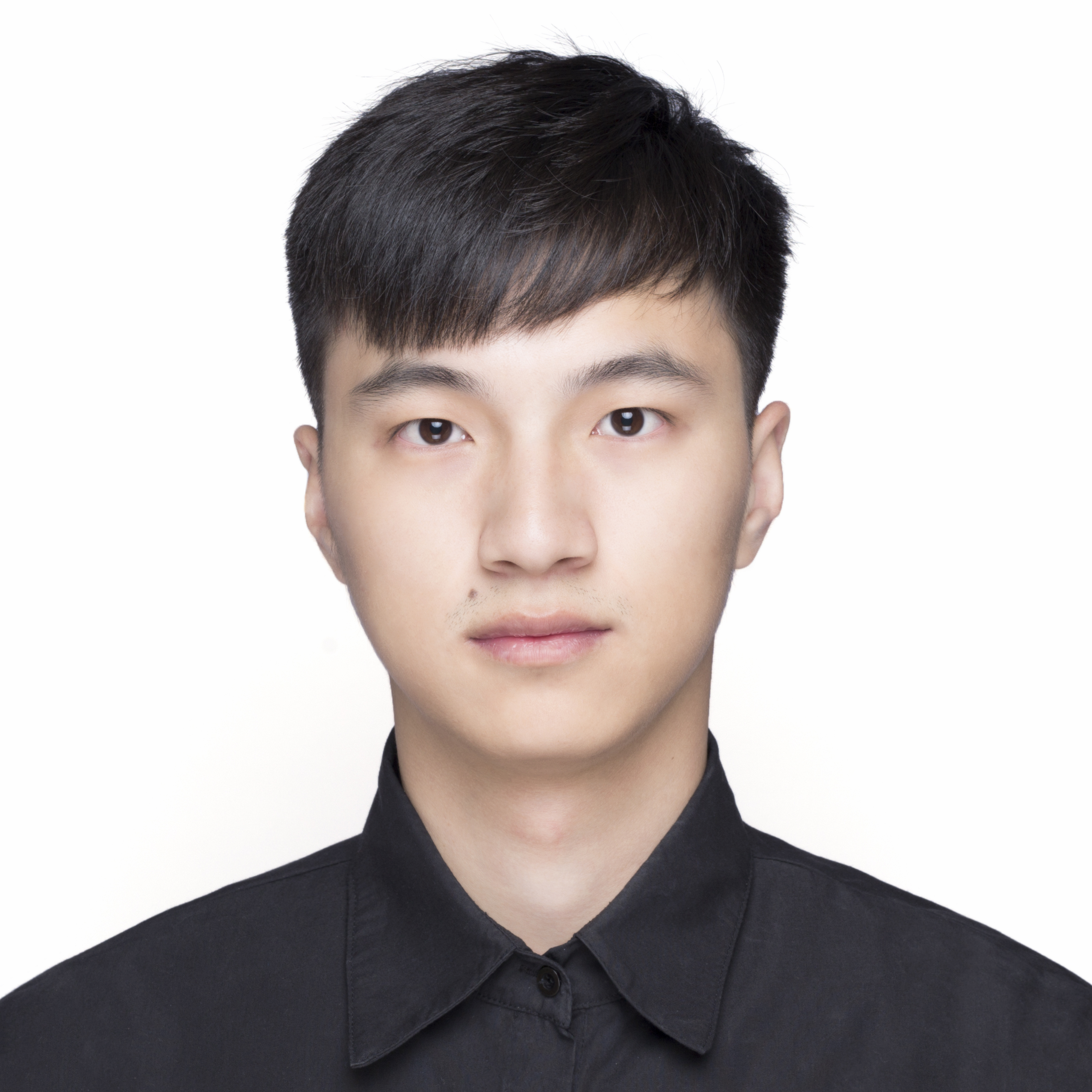}}]{Peng Qian}
received his BSc degree in software engineering from Yangtze University, MSc degree in computer science from Zhejiang Gongshang University, in 2018 and 2021. He is currently pursuing a Ph.D. at Zhejiang University. His research interests include blockchain security, graph neural network, and deep learning.
\end{IEEEbiography}

\vspace{-4em}
\begin{IEEEbiography}[{\includegraphics[width=1in,height=1.25in,clip,keepaspectratio]{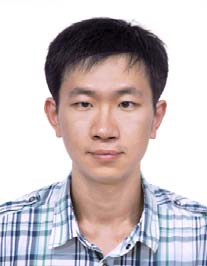}}]{Xiaoyang Wang}
received the BSc and MSc degrees in computer science from Northeastern University, China, in 2010 and 2012, respectively, and the PhD degree from the University of New South Wales, Australia, in 2016. He is a professor in Zhejiang Gongshang University, Hangzhou, China. His research interest includes query processing on massive graph data.
\end{IEEEbiography}

\vspace{-4em}
\begin{IEEEbiography}[{\includegraphics[width=1in,height=1.25in,clip,keepaspectratio]{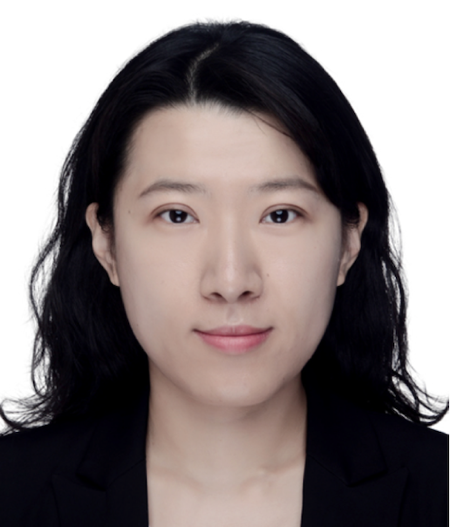}}]{Yuan Zhuang}
received her PhD from the College of Computer Science and Technology (CCST), Jilin University, China. Her research interests include blockchain security, machine learning, big data processing and distributed computing
\end{IEEEbiography}

\vspace{-4em}
\begin{IEEEbiography}[{\includegraphics[width=1in,height=1.25in,clip,keepaspectratio]{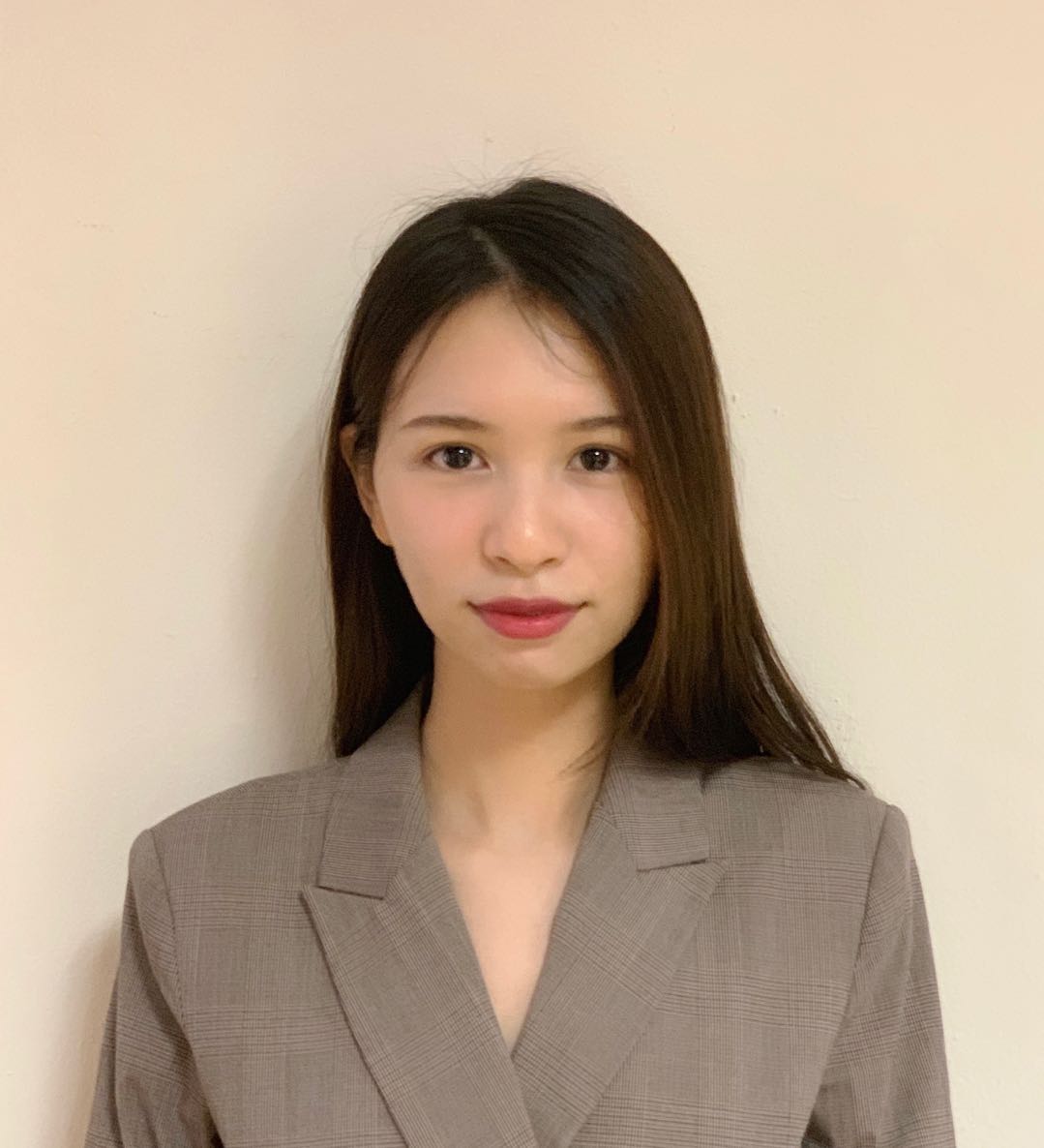}}]{Lin Qiu}
 is a PhD candidate at the Department
of Information Systems and Analytics, National University of Singapore, Singapore. Before
that, she obtained her bachelor degree from Xiamen University, China. Her research interests lie in deep learning, healthcare, and blockchain.
\end{IEEEbiography}

\vspace{-4em}
\begin{IEEEbiography}[{\includegraphics[width=1in,height=1.25in,clip,keepaspectratio]{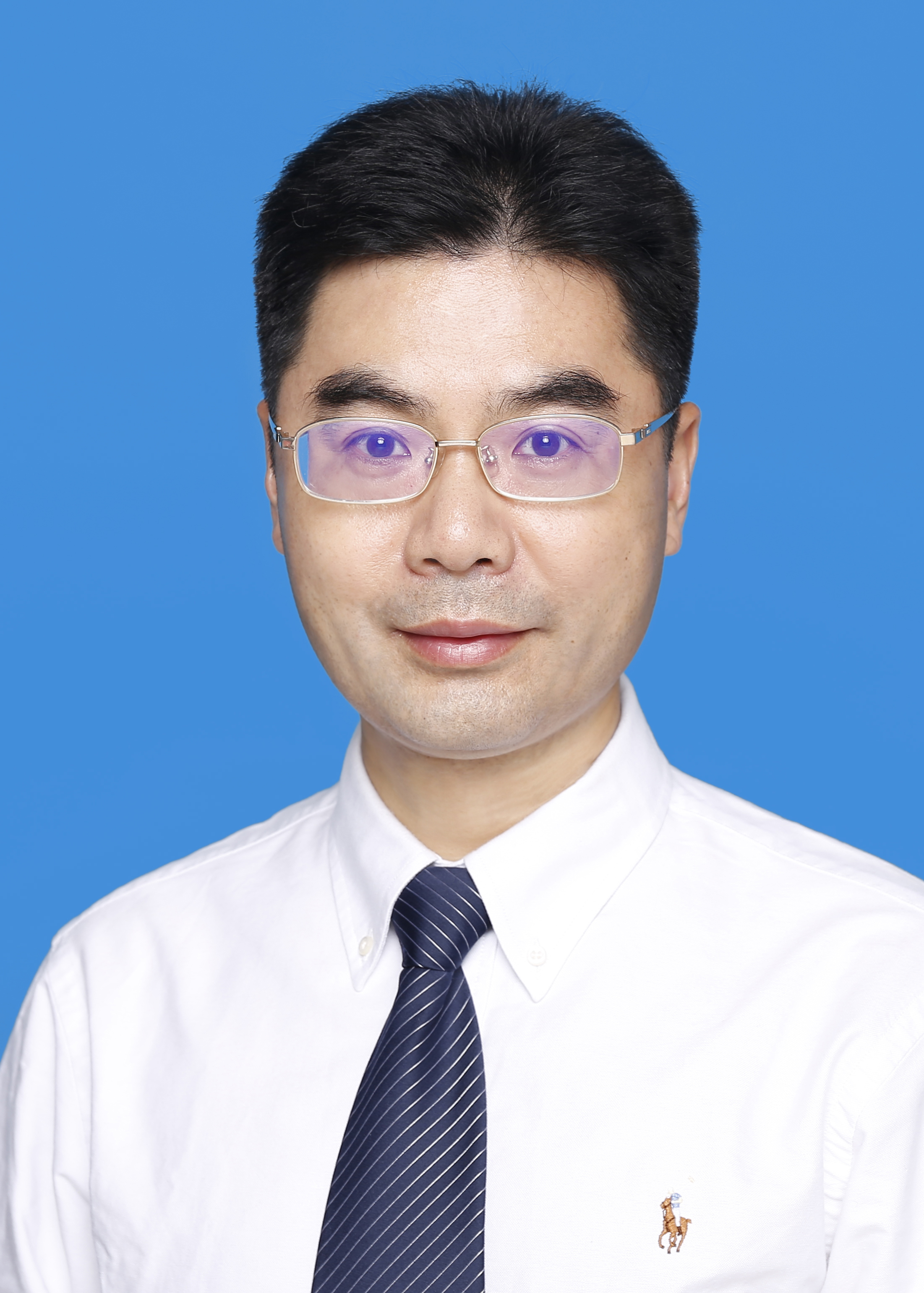}}]{Xun Wang}
is currently a professor at the School of Computer Science and Information Engineering, Zhejiang Gongshang University, China. He received his BSc in mechanics, Ph.D. degrees in computer science, all from Zhejiang University, China, in 1990 and 2006, respectively. His  research interests include intelligent information processing  and computer vision. He has published over 100 papers in high-quality journals and conferences. He is a member of the IEEE and ACM, and a distinguished member of CCF.
\end{IEEEbiography}

\fi

\end{document}